\documentclass{IEEEtran}

\usepackage{graphicx,epstopdf}
\usepackage{color}
\usepackage{multirow}
\usepackage{makecell}
\usepackage{amsmath,amssymb}
\usepackage{cases}
\usepackage{url}
\usepackage{color, soul}
\usepackage[normalem]{ulem}
\usepackage[sort,compress]{cite}
\usepackage{balance}
\usepackage{multirow}
\usepackage{rotating}
\usepackage{dsfont}
\usepackage{amssymb}
\usepackage{amsmath}

\usepackage[small,bf]{caption}
\usepackage[font=footnotesize]{subfig}

\newcommand{\singlesize}{0.5}

\usepackage{algorithm}
\usepackage{algpseudocode}

\usepackage{amsthm}
\newtheoremstyle{mythm}
{\topsep}   
{\topsep}   
{\itshape}      
{0pt}       
{\bfseries} 
{:}         
{5pt plus 1pt minus 1pt}    
{\thmname{#1}\thmnumber{ #2}\thmnote{ (#3)}}
\theoremstyle{mythm}

\newtheorem{definition}{Definition}

\newtheorem{proposition}{Proposition}
\renewcommand{\algorithmicrequire}{\textbf{Input:}}
\renewcommand{\algorithmicensure}{\textbf{Output:}}

\begin{document}
\title{Robust Transmissions in Wireless Powered Multi-Relay Networks with Chance Interference Constraints}
\author{
\IEEEauthorblockN{Jing Xu, \textit{Member, IEEE}, Yuze Zou, Shimin Gong, \textit{Member, IEEE}, Lin Gao, \textit{Member, IEEE}, Dusit Niyato, \textit{Fellow, IEEE}, and Wenqing Cheng}, \textit{Member, IEEE}
\thanks{Jing Xu, Yuze Zou, and Wenqing Cheng are with the School of Electronics Information and Communications, Huazhong University of Science and Technology, China. Email: \{xujing, zouyuze, chengwq\}@hust.edu.cn.}
\thanks{Shimin Gong is with the Shenzhen Institutes of Advanced Technology, Chinese Academy of Sciences, Shenzhen China. Email: sm.gong@siat.ac.cn.}
\thanks{Lin Gao is with the Department of Electronic and Information Engineering, Harbin Institute of Technology (Shenzhen), China. Email: gaol@hit.edu.cn.}
\thanks{Dusit Niyato is with the School of Computer Science and Engineering, Nanyang Technological University, Singapore. Email: dniyato@ntu.edu.sg.}
}

\maketitle

\begin{abstract}
In this paper, we consider a wireless powered multi-relay network in which a multi-antenna hybrid access point underlaying a cellular system transmits information to distant receivers. Multiple relays capable of energy harvesting are deployed in the network to assist the information transmission. The hybrid access point can wirelessly supply energy to the relays, achieving multi-user gains from signal and energy cooperation. We propose a joint optimization for signal beamforming of the hybrid access point as well as wireless energy harvesting and collaborative beamforming strategies of the relays. The objective is to maximize network throughput subject to probabilistic interference constraints at the cellular user equipment. We formulate the throughput maximization with both the time-switching and power-splitting schemes, which impose very different couplings between the operating parameters for wireless power and information transfer. Although the optimization problems are inherently non-convex, they share similar structural properties that can be leveraged for efficient algorithm design. In particular, by exploiting monotonicity in the throughput, we maximize it iteratively via customized polyblock approximation with reduced complexity. The numerical results show that the proposed algorithms can achieve close to optimal performance in terms of the energy efficiency and throughput.
\end{abstract}


\section{Introduction}
The next generation communications systems are anticipated to connect billions of devices arising from the popularity of wearable electronics and handhold devices as well as to provide orders of magnitude increase in capacity~\cite{5gsurvey17}. We envision that the traffic proliferation can be categorized into foreground and background communications, depending on human involvement in the communication process. The foreground communication involves human on one end, demanding agile responses with specific quality of service provisioning. Furthermore, we embrace a dramatic traffic increase in device-to-device (D2D) communications in the background without human intervention. With a number of autonomous and independent devices, the background communication will affect more significantly the overall network performance and require a strategic shift in the design of future wireless networks~\cite{densetut16}. Without human intervention, the energy supply firstly becomes a critical issue for network scalability and sustainability. It becomes impractical and costly to recharge or replace batteries for billions of D2D user devices. The increasing number of user devices also leads to a crowded usage of the spectrum resource. A spectral-efficient solution is to allow spectrum sharing with the existing licensed/primary wireless system such as the cellular network in an underlay manner~\cite{dstd17}. This requires the D2D user devices to precisely control their transmit power such that the interference to the cellular receiver is kept below pre-defined threshold.
\IEEEpubidadjcol

As one of the promising techniques, wireless power transfer provides a cost-effective way to sustain wireless communications, e.g.,~\cite{green17} and~\cite{luxiao}. Wireless information and power transfer can be implemented in either a time-switching (TS) or power-splitting (PS) scheme~\cite{zhangrui13}. The TS scheme divides the whole time slot into two sub-slots. One sub-slot is designated as the EH time, reserved for wireless energy transfer, while the other sub-slot is used for information transmission. The PS scheme splits a fraction (namely, the PS ratio) of the received signal power to energy harvester and feeds the other fraction to information receiver. In either the TS or PS scheme, the EH-enabled user devices have to alternate between the states of energy harvester and information receiver, incurring the study on rate-energy tradeoff~\cite{ratene13}. The TS and PS schemes have been extended to relay networks, in which the source nodes transfer power to the relays and the relays in return assist the information transmission, introducing the energy and information cooperation at the relays. \emph{Single relay models} in both the TS and PS schemes have been proposed in~\cite{relay13}, focusing on an amplify-and-forward (AF) relaying protocol. The authors in~\cite{ding16} and~\cite{select15} investigated the relay's scheduling policies in a multi-access model, in which multiple transceiver pairs are assisted by one EH relay with or without energy cooperation, respectively. \emph{Multi-antenna relay} has been studied in~\cite{liu17} and~\cite{artificial}. The authors in~\cite{liu17} focused on the classic three-node EH relay model, in which a multi-antenna relay operates in the decode-and-forward (DF) protocol. To maximize the end-to-end throughput, the PS protocol is implemented by an antenna clustering scheme that divides multiple antennas of the relay into two disjoint groups, i.e., one group for information decoding and the other for energy harvesting. The authors in~\cite{artificial} considered the existence of an eavesdropper that may intercept the information from the source node. The multi-antenna EH relay has to optimize its signal beamforming strategy to prevent information leakage to the eavesdropper, by generating Gaussian artificial noise (AN) signals. \emph{Multi-relay selection} for EH-enabled two-way communications is investigated in~\cite{twowayps} to maximize the data rate by jointly optimizing the set of relays, the relays' PS ratios and transmit power levels. An energy-threshold based multi-relay selection scheme is proposed in~\cite{mselect18} for each UE to decide locally whether to operate in EH or information forwarding mode. 

When multiple user devices are within one hop D2D communication range. They can join cooperative transmissions via short-range wireless communications. This provides the potential benefits of improved link quality and coverage, increased spectral and energy efficiency, reduced interference and power consumptions~\cite{densed2d16}. This motivates us to employ EH-enabled relays in end-to-end information delivery leveraging the relays' cooperation in both signal transmission and energy harvesting~\cite{coop15}. The energy densities at different relays can be different because of the time-varying channel conditions and the non-homogeneous distribution of the RF energy over space. By designing proper beamforming strategy of the power emitter, we can reshape the energy distribution over space and thus adjust the power transfer to different relays according to their power demands. By multi-users' signal cooperation, multiple relays can form a virtual multiple-input and multiple-output (MIMO) system and improve the throughput performance between a distant transceiver pair significantly while causing insignificant increase in the end-to-end delay, e.g.,~\cite{sinrelay2} and~\cite{virtualmimo}.

\emph{Multi-relays' collaborative beamforming} has been studied to make use of the harvested energy at multiple user devices. The authors in~\cite{twobf17} considered two user devices wirelessly powered by an access point in the downlink. Two cooperation schemes were proposed to maximize the sum rate in the uplink, by joint optimization of the time allocation, energy beamforming, and power allocation strategies. The authors in~\cite{liu2016wireless} optimized the PS ratios of multiple relays to maximize the data rate between a single-antenna transceiver pair. The maximization is decomposed into multiple local problems at individual relays, in which each relay only needs to optimize its own PS ratio. In our previous work~\cite{twc2016}, with fixed EH rates at individual relays, we optimized the power amplifier coefficient at each relay in the TS scheme. The authors in~\cite{Globecom16:Gong} considered channel-dependent EH rates at individual relays, which are controllable by the energy beamforming strategy. A joint optimization of collaborative relay beamforming and the energy beamforming strategies are proposed in the TS scheme to maximize the throughput of a distant transceiver pair. Considering a total power budget constraint for all the EH relays, the authors in \cite{netbf_sumpower} designed the collaborative beamforming and power allocation strategies to maximize the average throughput over consecutive time slots. However, the aforementioned works mainly focus on parts of the transmission control problems in wireless powered networks, considering either the TS or PS protocol, while overlooking the strong couplings among wireless power transfer, energy harvesting, and transmit power control strategies. In this paper, we provide a unified framework for throughput maximization under both the TS and PS protocols. The optimization builds a thorough connection between the throughput performance and different operating parameters, related to wireless power transfer, energy harvesting, and transmit power control. The parameters for wireless power transfer and energy harvesting control the energy supply to different relays. The transmit power control determines the energy demand at individual relays. A well balanced energy supply and demand will maximize the multi-relay assisted transmission performance.

In particular, we consider a wireless powered multi-relay network composed of a hybrid access point (HAP) transmitting information to a distant receiver and sharing the same spectrum with a underlaying cellular system. The transmit performance is improved by multiple relays wirelessly powered by the HAP's energy beamforming. Given the relays' power demands, the optimal beamforming strategy has been revealed in~\cite{xujie15} to maximize the energy transfer to the relays. It has been shown that it is generally not optimal for all the relays to transmit with their peak power, due to the relays' different channel conditions. The reason is that every relay has two effects on the performance of information transmission. On one hand, it helps the transmission by forwarding the information. On the other hand, it harms the transmission by also forwarding the noise signal. For some relay with bad channel conditions, it will experience high noise power and cause performance degradation at the receiver if the relay transmits with maximum power. This implies that the relays' power demands can differ from each other and depend on the relays' channel conditions. Hence, some relays are \emph{energy-starving} while the other relays can be \emph{energy-abundant}. The energy-starving relays become the bottleneck for the transmission from the HAP to the targeted receiver. With multiple EH relays, the relays' transmission control becomes more complicated. This motivates a joint optimization of the HAP's energy beamforming as well as the relays' EH and power control strategies, subject to interference constraints at cellular receivers. The main contributions of this paper are summarized as follows.
\begin{itemize}
  \item Unified Throughput Optimization with the TS and PS Schemes: The optimization builds a thorough connection between the throughput performance and different operating parameters, related to wireless power transfer, energy harvesting, and transmit power control. The parameters for wireless power transfer and energy harvesting control the energy supply to different relays. The power control determines the energy demand at individual relays.
  \item Multi-relay Robust Transmission Strategies via Monotonic Optimization: The coupling of design variables and the probabilistic interference constraints make the proposed robust optimization problem non-convex. By examining monotonicity of the objective and verifying normal feasible regions, we can achieve enhanced throughput throughput by successive polyblock approximation. To the best of our knowledge, we are the first to present an optimization for a wireless powered multi-relay network that is close to optimal and robust to channel uncertainties.
  \item Energy Efficiency Comparison between the TS and PS schemes: In the PS scheme, individual relay can adjust its PS ratio to match power demand and supply as well as information transfer. On the contrary, the EH time in the TS scheme is controlled by the HAP and all UEs have the same EH time. With the robust formulation, we show that the PS scheme is more flexible in the transmission control and generally achieves higher energy efficiency. However, when the HAP's transmit power is low, such flexibility becomes trivial and results in degraded performance compared to that of the TS scheme.
\end{itemize}

The rest of the paper is organized as follows. We present the EH-enabled multi-relay network model in Section~\ref{sec:sys-mod}. In Section~\ref{sec_ts} and Section~\ref{sec_ps}, we solve the throughput maximization problems subject to probabilistic interference constraints, under the TS and PS schemes, respectively, by developing a unified optimization framework. Numerical results and conclusions are given in section~\ref{sec:simulation} and~\ref{sec:conc}, respectively.

\begin{table}[t]
\centering\caption{Table of Notations}\label{tab:notations}
\begin{tabular}{|c|p{6.5cm}|}
\hline
{\bf Notation} & {\bf Description} \\ \hline
$K$				&	The number of antennas at HAP		\\ \hline
$\mathcal{N}$	& 	Set of $N$ relays, i.e., $\mathcal{N} = \{1,2,\ldots,N\}$ \\ \hline
$\mathcal{C}$	&	Set of $M$ CUEs, i.e., $\mathcal{C} = \{1,2,\ldots,M\}$		 \\ \hline
$\eta$ 	& 			RF power conversion efficiency	\\\hline
$\zeta$ & Limit of interference violation probabilities at CUEs \\ \hline
$p_n$	&	Normalized transmit power of relay-$n$ and denote $c_n = \sqrt{p_n}$ for ease of analysis	 \\\hline
$p_{\mathrm{o}}$, $s$		&	Normalized transmit power and complex information symbol with unit power of HAP \\\hline
$w $, $t$		&	EH and information transmission time in the TS scheme \\\hline
$\rho_n$, ${\bf \rho}$		& PS ratio of relay-$n$, and a vector of the relays' PS ratios $\boldsymbol{\rho}\triangleq [ \rho_1, \rho_2, \ldots, \rho_N ]^T$ \\\hline
${\bf w}_{\mathrm{e}}$, ${\bf w}_1$ &	HAP's beamforming vectors for power transfer and information transmission in the first hop\\\hline
${\bf w}_{\mathrm{p}}$ 	&	HAP's beamforming vector for mixed power and information transfer in the first hop	\\\hline
${\bf f}_n$		&	Complex channel from the HAP to relay-$n$ in vector form ${\bf f}_n = [f_{1n},f_{2n},\ldots,f_{Kn}]^T$ \\
\hline
${\bf g}$		& 	Complex channel from $N$ relays to the receiver in vector form ${\bf g} = [g_{1},g_{2},\ldots,g_{N}]^T$	\\
\hline
${\bf z}_m$	   &	Complex channel from $N$ relays to CUE-$m$ in vector form ${\bf z}_m = [z_{1m},\ldots,z_{nm},\ldots,z_{Nm}]^T$ \\
\hline
$\left(\mathbf{u}_{m}, \mathbf{S}_{m}\right)$ & The 1st- and 2nd-order moment of channel ${\bf z}_m$\\ \hline
$\gamma$, $\gamma_d$		&	SNR at receiver without or with a direct link from the HAP to the receiver	 \\ \hline
$y_n$, ${\bf y}$ & The received information signals at relay-$n$, and its vector form ${\bf y} = [y_1, \ldots, y_n, \ldots, y_N]^T$ \\ \hline
$x_n$, ${\bf x}$ & Power amplifying coefficient of relay-$n$, and its vector form ${\bf x} = [x_1,\ldots,x_n,\ldots,x_N]^T$ \\ \hline
$\sigma_n$, $v_d$ & Noise signals at relay-$n$, and at the targeted receiver \\ \hline
$\text{D}(\cdot)$ & Diagonal matrix with the specified diagonal element \\ \hline
${\bf \Delta}(\cdot)$ & Diagonal matrix by setting off-diagonal elements to zeros \\\hline
\end{tabular}
\end{table}

\section{System Model}\label{sec:sys-mod}

\begin{figure}[t]
\centering
\includegraphics[width=0.5\textwidth]{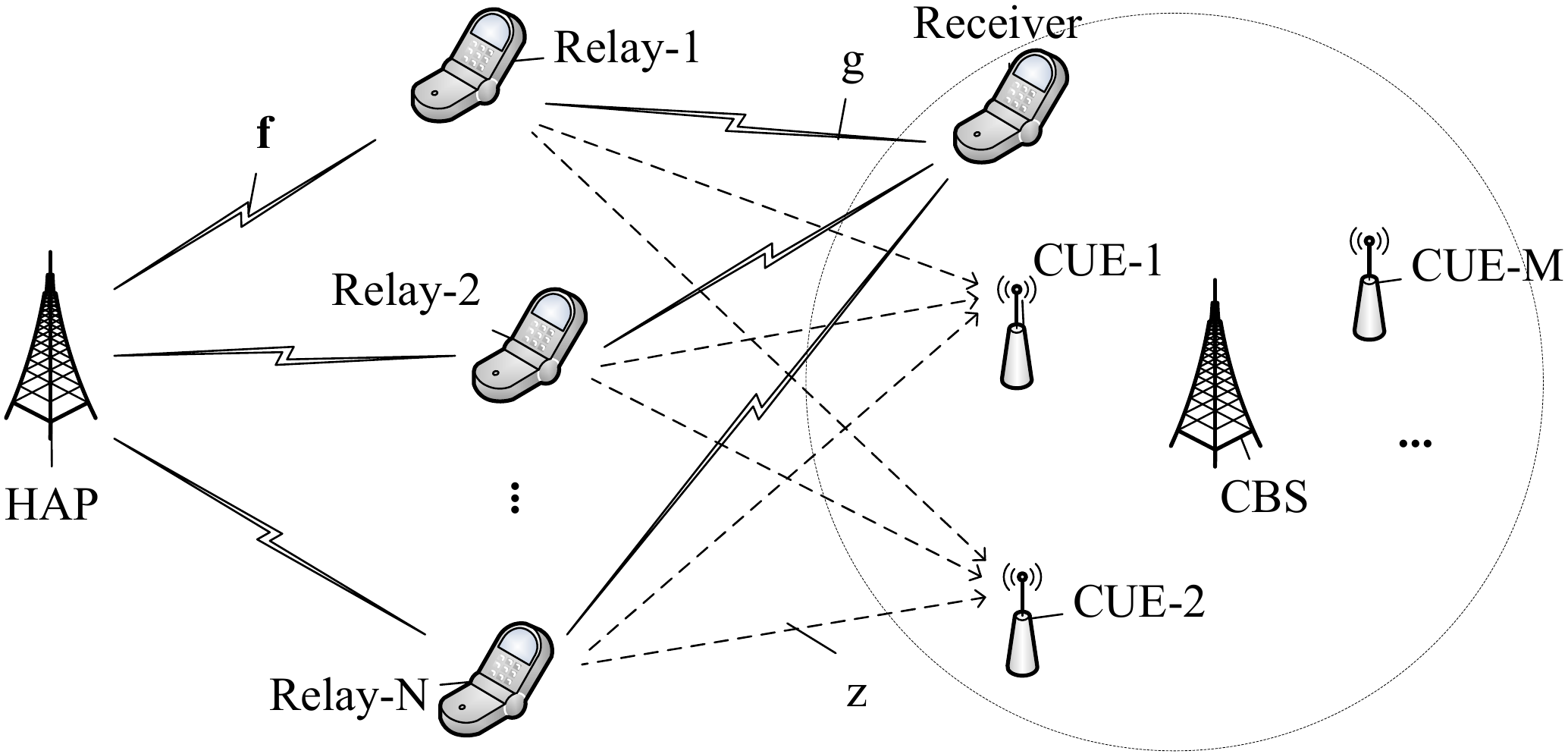}
\caption{Wireless powered multi-relay assisted two-hop transmissions.}\label{fig:system-model}
\end{figure}

We consider a wireless powered multi-relay network consisting of an HAP and multiple relays underlaying a downlink cellular system, e.g., the HAP and D2D user devices can be deployed on the edge of the cellular coverage and share the same spectrum with the cellular receiver. The goal is to maximize throughput from the HAP to a certain receiver. The HAP has $K$ antennas and aims to transmit information to a distance receiver with the help of relays. The HAP serves as a power and information source for the relays using either the TS or PS scheme~\cite{relay13}. All the relays are equipped with a single antenna and capable of harvesting RF energy from the HAP. Through dedicated energy beamforming, the HAP can control the rates of information and power transfer to the relays. The HAP also acts as a central controller governing information transmission by the relays and coordinating with the cellular system. With multiple receivers, the HAP schedules the information transmission for each of them by using time division multiple access (TDMA). Specifically, each receiver is registered at the HAP and assigned a fixed time slot with unit time length. Table~\ref{tab:notations} lists the major notations used in the paper. Fig.~\ref{fig:system-model} shows the network structure.


To improve the transmit performance from the HAP to a receiver, we adopt cooperative relay beamforming in the amplify-and-forward (AF) protocol. The set of relays is denoted by $\mathcal{N}=\{1,2,\ldots,N\}$. Due to the reuse of licensed spectrum in an underlay manner, transmission of the relays may cause interference to the cellular user equipments (CUEs), the set of which is denoted by $\mathcal{C}\triangleq\{1,2,\ldots,M\}$. Nevertheless, it is acceptable if the aggregate interference to each of the CUEs is less than a pre-defined threshold. The downlink transmissions from the cellular base station to the CUEs can also cause interference to the underlaid relays. In practice, through channel estimation, the power level of the noise and the cellular interference to each relay can be known in advance and viewed as a constant~\cite{gsm17}. For example, each relay can estimate the power level in a silence period before initiating information transmissions.

\subsection{Two-hop Wireless Powered Multi-Relay Transmission}

The information transmission follows a two-hop half-duplex relay protocol. In the first hop, the HAP beamforms the information signals to the relays with a fixed transmit power. This fixed power is reasonable in that the HAP can coordinate with the cellular base station. Meanwhile, the relays harvest energy from the information signals by the TS or PS scheme. Then, in the second hop, the relays use the harvested energy to amplify and forward the received signals to the targeted receiver. Let ${\bf f}_n \triangleq \left[f_{1n}, \ldots, f_{kn}, \ldots, f_{Kn}\right]^T\in\mathbb{C}^K$ denote the complex channel vector from the HAP to relay-$n$. Let ${\bf g}\triangleq[g_1,\ldots, g_n,\ldots, g_N]^T\in\mathbb{C}^N$ denote the channel vector from $N$ relays to the receiver. Likewise, let ${\bf z}_m \triangleq [z_{1m},\ldots, z_{nm}, \ldots, z_{Nm}]^T\in\mathbb{C}^N$ denote the channel vector from $N$ relays to CUE-$m$, for $m\in\mathcal{C}$. All channels exhibit block fading. We focus on a simple case in which the direct link from the HAP to the receiver is not available, e.g., due to blockage or undesirable signal quality. A similar model has been considered in many existing works, e.g.,~\cite{twc2016} and~\cite{nodirect14}. 

\begin{figure}[t]
\centering
\subfloat[Time division in the TS scheme.]{\includegraphics[width=0.5\textwidth]{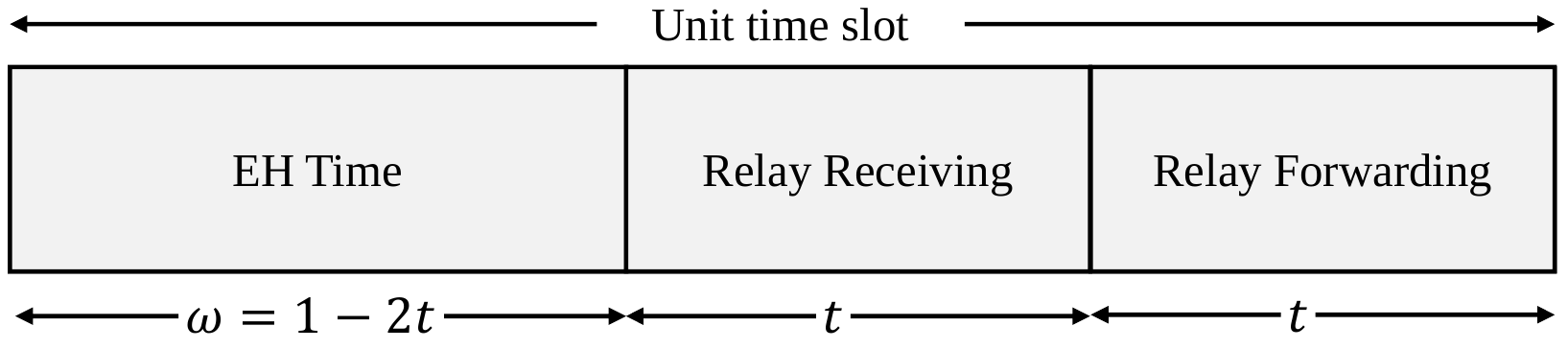}\label{fig:EH-TS}}\\
\subfloat[Power division in the PS scheme.]{\includegraphics[width=0.5\textwidth]{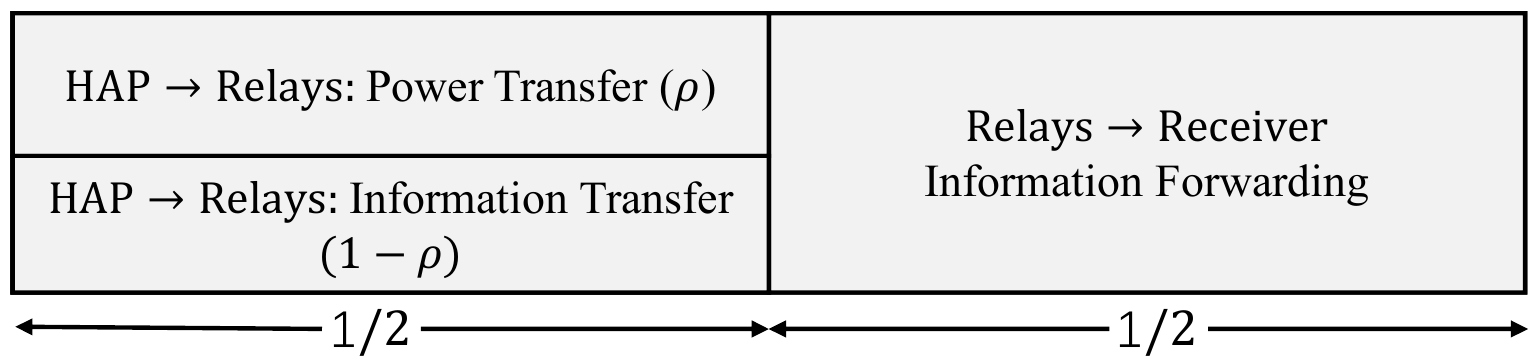}\label{fig:EH-PS}}
\caption{The TS and PS schemes for information and power transfer.}
\label{fig:EH}
\end{figure}

Let $m_n \triangleq y_n + \sigma_n$ denote the signal received by the relay-$n$, where $y_n$ denotes the received information signal transmitted from the HAP, and $\sigma_n\sim \mathcal{CN}(0,1)$ represents the complex Gaussian noise with zero mean and unit variance. Here, by assuming unit noise power, we normalize all the transmit powers by the noise power to simplify the problem formulation. Furthermore, we also normalize the amount of energy harvested from the received signals to the absolute noise power $N_o$. Hence, the normalized power of the received signal $m_n$ is given by $1 + |y_n|^2$. Accordingly, we define the power amplifying coefficient of relay-$n$ as follows:
\begin{equation}
	\label{equ_amp}
	x_n = \left(\frac{p_n}{|y_n|^2 + 1}\right)^{1/2},
\end{equation}
where $p_n$ denotes the transmit power of relay-$n$, depending on the HAPs' beamforming strategy and the relay's EH parameters. At the receiver, the received signal is given by $u = \sum\nolimits_{n\in\mathcal{N}} x_n y_n g_n + \sum\nolimits_{n\in\mathcal{N}} x_n g_n\sigma_n + v_d$, where $v_d \sim \mathcal{CN}(0,1)$ denotes the normalized noise in the receiver. The first term, $\sum\nolimits_{n\in\mathcal{N}} x_n y_n g_n$, contains signals from the HAP and relays. The second term, $\sum\nolimits_{n\in\mathcal{N}} x_n y_n g_n$, is the forwarded noise by the relays. Thus, the signal-to-noise ratio (SNR) of the information at the receiver is given by
\begin{equation}
	\gamma(\mathbf{x}, \mathbf{y}) = \frac{\left| \sum_{n\in\mathcal{N}} x_n y_n g_n \right|^2}{1 + \sum_{n\in\mathcal{N}}|x_n g_n|^2 } = \frac{ |({\bf x}\circ{\bf y})^H{\bf g}|^2}{1+ {\bf x}^T \text{D}({\bf g}\circ{\bf g}){\bf x}}.
\label{equ:gamma}
\end{equation}
Here ${\bf f}^H_n$ represents the Hermitian transpose of vector ${\bf f}_n$. The symbol $\circ$ denotes the Hadamard product. $\text{D}({\bf g}\circ{\bf g})$ is a diagonal matrix with the diagonal elements specified by the vector ${\bf g}\circ{\bf g}$.

\subsection{Wireless Power and Information Transfer}

We consider two different schemes for the relays to harvest RF energy. In the TS scheme, a time slot is divided into two phases, as shown in Fig.~\ref{fig:EH}(a). The first phase with length $w$, i.e., the EH time, is used for the HAP to beamform wireless energy to the relays. The second phase is used for relay-assisted information transmission. It is divided further into two equal sub-phases for information receiving and forwarding by the relays to the receiver. Note that the lengths of the two sub-phases can also be optimized by guided one-dimension search~\cite{gsm18tcom}. Nonetheless, for ease of derivation and presentation, we assume the time length $t\in(0,1/2)$ to be identical in two sub-phases and the relays' EH time in the first phase is $w \triangleq 1-2t $. A similar model has been studied in~\cite{Globecom16:Gong} and~\cite{equalts16}. To achieve optimal throughput, the HAP can schedule the EH time $w$ as well as the channel time $t$ for information transmission/forwarding, given channel conditions and energy demands of the relays. Notice that all the relays have the same EH time controlled by the HAP.

Alternatively, in the PS scheme, the signal transmitted by the HAP is used for both EH and information receiving by the relays. Therefore, the entire time slot is divided into two phases, as shown in Fig.~\ref{fig:EH}(b). The first phase is for the HAP to transmit information and energy to the relays simultaneously, and the second phase is for the relays to forward the information to the receiver. Again, the lengths of these two phases can be optimized. Nonetheless, we assume that they are identical~\cite{zhao17ps}. At the relay, the ratio for extracting energy from the received signals is denoted by $\rho$, and thus the ratio of signal power for information reception is $1-\rho$. Unlike the TS scheme, each relay in the PS scheme can set different PS ratio $\rho$, to best match a beamforming strategy of the HAP and its energy demand.

\subsection{Channel Uncertainty and Interference Constraint}

We aim to maximize the throughput at the receiver, constrained by the interference limit to the CUEs and the amount of energy harvested by the relays. Note that the channels ${\bf f}_n$ and ${\bf g}$ in the first and second hops can be perfectly known by the HAP and relays, respectively. This is possible as they are all in the same system and channel estimation can be performed in a timely fashion. However, exact channel information may not be available to evaluate the interference to CUEs. As the CUEs operate on a different system, timely response from the CUEs may not be possible in practice. Therefore, we assume that the channel $\mathbf{z}_m$ from the relays to CUE-$m$ is uncertain, for $m\in\mathcal{C}$. Note that the moments of $\mathbf{z}_m$ are relatively easier to estimate with higher accuracy~\cite{Globecom16:Gong}. Hence, we characterize the probability density function (pdf) of $\mathbf{z}_m$ by its first- and second-order moments as follows:
\begin{equation}
	\mathbb{P}_{m} \in \mathcal{P}\left(\mathbf{u}_{m}, \mathbf{S}_{m}\right),
\end{equation}
where $\mathcal{P}(\mathbf{u}_{m}, \mathbf{S}_{m})$, or $\mathcal{P}_{m}$ for short, denotes the set of pdfs with the moments $\left(\mathbf{u}_{m}, \mathbf{S}_{m}\right)$. As such, we define the CUEs' interference constraints in a probabilistic form as follows:
\begin{equation}\label{con_irobust}
\max_{\mathbb{P}\in \mathcal{P}_m} \mathbb{P}( \phi_m({\bf p})\geq \bar{\phi}_m) \leq \zeta, \quad \forall \, m\in\mathcal{C}	 ,	
\end{equation}
where ${\bf p} \triangleq [p_1,\ldots, p_n,\ldots, p_N]^T$ denotes a vector of the relays' transmit power, and $\bar{\phi}_m$ represents the tolerance of CUE-$m$ to the aggregate interference $\phi_m({\bf p})  \triangleq \sum\nolimits_{n\in\mathcal{N}} |z_{nm}|^2 p_n$. We denote $\mathbb{P}(\phi_m({\bf p})\geq \bar{\phi}_m)$ as the ``interference violation probability'' at the CUE-$m$. Hence, the constraint in \eqref{con_irobust} requires that the worst-case interference violation probabilities have to be upper bounded by a probability limit $\zeta$. The worst case is with respect to all distributions in set $\mathcal{P}_{m}$ with the given moments $\left(\mathbf{u}_{m}, \mathbf{S}_{m}\right)$. Note that the chance constraints in~\eqref{con_irobust} only define the CUEs' interference requirements in the second hop when the relays transmit simultaneously. We omit the interference introduced by the HAP to the CUEs in the first hop, as the HAP is assumed to be deployed at the edge of cellular coverage and the direct link from HAP to each CUE is weak. Moreover, due to a shorter distance between D2D user devices and CUEs, the EH relays introduce much higher interference to CUEs than that introduced by the HAP. Hence, we consider that the interference constraint also holds in the first hop if \eqref{con_irobust} holds in the second hop.

\section{Robust Throughput Maximization with the TS Scheme}\label{sec_ts}

In the TS scheme, the HAP can control its beamforming strategies in two phases, i.e., energy transfer and information transmission. Let $p_{\mathrm{o}}$ denote transmit power of the HAP normalized to the noise power and $s\in\mathbb{C}$ be the complex information symbol with unit power delivered from the HAP to the receiver. The transmit power $p_{\mathrm{o}}$ can be optimized by one-dimension search as discussed in~\cite{twc2016}. Without loss of generality, we assume that $p_{\mathrm{o}}$ is fixed in this paper, e.g., we can set $p_{\mathrm{o}}$ as the maximum transmit power of the HAP.

Define ${\bf w}_{\mathrm{e}}$ as normalized beamforming vector of the HAP during the EH phase. Then, the energy signal transmitted by the HAP is ${\bf x}_s = \sqrt{p_{\mathrm{o}}}{\bf w}_{\mathrm{e}} s$. By varying the beamformer ${\bf w}_{\mathrm{e}}$, we can control the HAP's transmit power and adjust the power transfer to different relays according to their channel conditions and EH capabilities. Then, the relays' power budget constraints are given as follows:
\begin{equation}\label{equ_powerts}
	p_n \leq \bar{p}_n(t,{\bf w}_{\mathrm{e}}) \triangleq \eta p_{\mathrm{o}} {\bf f}^H_n {\bf W}_{\mathrm{e}} {\bf f}_n(1 - 2t)/t, \quad \forall n \in \mathcal{N}.
\end{equation}
We denote $\eta$ as the energy conversion efficiency, which can also account for the energy consumption in the circuit and information reception. The matrix variable ${\bf W}_{\mathrm{e}} \triangleq {\bf w}_{\mathrm{e}} {\bf w}_{\mathrm{e}}^H$ denotes the transmit covariance of the HAP. Note that we can relax the equality by ${\bf W}_{\mathrm{e}} \succeq {\bf w}_{\mathrm{e}}{\bf w}_{\mathrm{e}}^H$ and adopt ${\bf W}_{\mathrm{e}}$ to be the design variable. Then, once we determine the optimal ${\bf W}_{\mathrm{e}}^*$, we can retrieve the energy beamformer ${\bf w}_{\mathrm{e}}$ by eigen-decomposition if ${\bf W}_{\mathrm{e}}^*$ is rank-one. Otherwise, we can construct a stochastic beamformer ${\bf w}_{\mathrm{e}}(t)$ such that ${\bf W}_{\mathrm{e}} = \mathbb{E}[{\bf w}_{\mathrm{e}}(t){\bf w}_{\mathrm{e}}^H(t)]$~\cite{xxwu}. In the power budget constraint~\eqref{equ_powerts}, we actually have assumed a linear EH model with a constant power conversion efficiency $\eta$, i.e., the total harvested energy takes up a fixed portion of RF power at the receiving antenna. A more practical nonlinear EH model has been presented in~\cite{noneh}, in which $\eta$ is measured to be a function of the RF power at the receiving antenna. Our subsequent analysis will show that our problem formulation and solution can be extended to the non-linear EH model.

Let ${\bf w}_1$ be the HAP's beamforming strategy for information transmission in the first hop. Then, the signal received at relay-$n$ and its power amplifying coefficient are given by
\[
m_n = \sqrt{p_{\mathrm{o}}}{\bf f}_n^H{\bf w}_1 s + \sigma_n \text{ and } x_n = \left(\frac{p_n}{1+ p_{\mathrm{o}} {\bf f}^H_n {\bf W}_1 {\bf f}_n}\right)^{1/2},
\]
respectively, where ${\bf W}_1 = {\bf w}_1 {\bf w}_1^H$ denote the transmit covariance of the HAP in the first hop.

\subsection{Throughput Maximization Problem}

We aim to maximize the throughput in the TS scheme~\eqref{equ:obj1}, given the chance constraint~\eqref{equ:inter-constr1} that the interference to CUEs is higher than a threshold is maintained below a certain probability and the power budget constraint~\eqref{equ:power-constr1}. The decision variables are the channel time $t$ for information transmission, the relays' transmit power vector ${\bf p}$, and the HAP's beamforming strategies $({\bf W}_e, {\bf W}_1)$ in two phases. The throughput maximization problem is given as follows:
\begin{subequations}\label{equ:original-form1}
\begin{align}
	\max_{t, \mathbf{c},{\bf W}_{\mathrm{e}},{\bf W}_1}~~ & t \log\left(1 + \frac{{\bf c}^T {\bf H}{\bf G}{\bf c}}{1 + {\bf c}^T {\bf B} {\bf c}}\right) \label{equ:obj1}\\
\text{s.t.}~~ & \max_{\mathbb{P}\in \mathcal{P}_{m} } \mathbb{P} \left(\phi_m(\mathbf{c}) \geq \bar{\phi}_m \right) \leq \zeta, \quad \forall m \in\mathcal{C}, \label{equ:inter-constr1}\\
& p_n t \leq (1 - 2t) \eta p_{\mathrm{o}} {\bf f}^H_n {\bf W}_{\mathrm{e}} {\bf f}_n, \quad \forall n \in \mathcal{N},\label{equ:power-constr1}	\\
& t \in [0, 1/2].\label{equ:time-constr1}
\end{align}
\end{subequations}
Here we denote ${\bf c} = [c_1,\ldots, c_n,\ldots, c_N]^T$ where $c_n \triangleq \sqrt{p_n}$ is an alias for the transmit power of relay-$n$ for ease of analysis. We also define the following notations for convenience:
\begin{align*}
& {\bf H} = \text{D}\left(\left[\frac{|h_1|^2}{1 + p_{\mathrm{o}} |h_1|^2}, \frac{|h_2|^2}{1 + p_{\mathrm{o}} |h_2|^2},\ldots,\frac{|h_N|^2}{1 + p_{\mathrm{o}} |h_N|^2}\right]\right),\\
& {\bf B} = \text{D}\left(\left[\frac{|g_1|^2}{1+p_{\mathrm{o}} |h_1|^2},\frac{|g_2|^2}{1+p_{\mathrm{o}} |h_2|^2}, \ldots, \frac{|g_N|^2}{1+p_{\mathrm{o}} |h_N|^2}\right]\right),\\
& {\bf G} = {\bf g}{\bf g}^H \text{ and } h_n = {\bf f}^H_n {\bf w}_1 \text{ for } n\in\{1,2,\ldots,N\},
\end{align*}
Problem~\eqref{equ:original-form1} maximizes the per-time-slot throughput performance, without considering the energy accumulation over different time slots. Typically, the harvested energy at each relay can be stored in a super-capacitor, which is suitable for energy harvesting from RF signals as it has much larger recharge cycles (theoretically infinite and more than one million practically), high charge and discharge efficiency at small energy levels~\cite{luxiao}. However, due to the super-capacitors' higher self-discharge rate, the stored energy is leaking out at a faster rate than the conventional rechargeable batteries. Hence, from a practical viewpoint, we assume that all the leftover energy will be depleted or leaking out before wireless power transfer in next time slot. Such harvest-use model without energy accumulation has also been studied in~\cite{twc2016}.

It is clear that obtaining an optimal solution to problem~\eqref{equ:original-form1} is not straightforward due to non-convex couplings of the decision variables at different relays. A close inspection reveals that the energy beamformer ${\bf W}_{\mathrm{e}}$ relates to the power budget and interference constraints in \eqref{equ:inter-constr1} and \eqref{equ:power-constr1}, respectively. Likewise, the information beamformer ${\bf W}_1$ only relates to the SNR performance in \eqref{equ:obj1}. This observation motivates us to update ${\bf W}_{\mathrm{e}}$ and ${\bf W}_1$ iteratively toward the solution. Specifically, given a solution $(t,{\bf c},{\bf W}_{\mathrm{e}})$ to problem~\eqref{equ:original-form1}, the power budget $\bar p_n(t,{\bf w}_{\mathrm{e}})$ at each relay is fixed and known. As such, the information beamformer ${\bf W}_1$ can be updated in a conventional beamforming problem similar to~\cite{liang2011}. The update of ${\bf W}_1$ leads to a new round optimization on $(t,{\bf c},{\bf W}_{\mathrm{e}})$. Note that either the update of ${\bf W}_{\mathrm{e}}$ or ${\bf W}_1$ improves the throughput performance. Such iterative update will terminate when no further improvement can be achieved. However, the main difficulty lies in that the problem in~\eqref{equ:original-form1} is non-convex and challenging to solve even with fixed ${\bf W}_1$.

\subsection{Problem Reformulation and Structural Properties}

In the sequel, we explore the structural property of the problem in \eqref{equ:original-form1} with fixed information beamforming ${\bf W}_1$. By a change of variable, we can view SNR $\gamma$ in \eqref{equ:gamma} as the decision variable and represent the objective function in~\eqref{equ:obj1} in a simpler form $r(t,\gamma)=t\log(1+\gamma)$. Note that the new objective is monotonically increasing in both $t$ and $\gamma$. This implies that its optimum will be achieved on the boundary of its feasible region. This observation motivates us to apply the \emph{monotonic optimization algorithm} by further exploring the structural properties of the feasible region of $(t,\gamma)$, which can be represented as follows:
\begin{equation}\label{set_omega1}
	\Omega \,=\,\left\{ {\bf z} \triangleq(t,\gamma) \left| 0\leq\gamma\leq \frac{{\bf c}^T{\bf A} {\bf c}}{1+{\bf c}^T {\bf B}{\bf c}}, \eqref{equ:inter-constr1}-\eqref{equ:time-constr1} \right. \right\},
\end{equation}
where ${\bf A} \triangleq {\bf H}{\bf G}$ and ${\bf B}$ depend on the beamforming strategy of the HAP and channel conditions of the relays. Hence, we can reformulate the problem~\eqref{equ:original-form1} in a compact form as follows:
\begin{equation}\label{prob_simple_r}
\max\{t\log(1+\gamma): (t,\gamma)\in\Omega\}.
\end{equation}

The theory of monotonic optimization was developed in~\cite{tuy} and has been applied to wireless communications~\cite{monot2}. By leveraging the monotonicity in the objective function, the monotonic optimization algorithm can achieve convergence to the global optimum for a set of non-convex problems if the feasible region is a \emph{normal set}, which is defined as follows:
\begin{definition}[\emph{Normal set}]\label{def_normal}
Let $\Omega$ be a compact set. For any ${\bf z} \in \Omega$, if ${\bf z}'\in\Omega$ for all ${\bf z}'\preceq {\bf z}$, we say that $\Omega$ is normal.\footnote{${\bf z}'\preceq {\bf z}$ (or ${\bf z}'\succeq {\bf z}$) denotes that ${\bf z} - {\bf z}'$ is piecewise non-negative (or non-positive).}
\end{definition}
\begin{definition}[\emph{Upper boundary}]\label{def_upbd}
Let ${\bf z}\in\Omega$ and $\Omega$ is normal. For any ${\bf z}{'}\succeq{\bf z}$ and ${\bf z}{'}\neq{\bf z}$, if ${\bf z}{'}\notin \Omega$ then ${\bf z}$ is an upper boundary point. The set of all upper boundary points is the upper boundary of $\Omega$, denoted by $\bar{\Omega}$.
\end{definition}
Note that a normal set $\Omega$ can be non-convex and difficult to characterize. For ease of analysis, we can approximate it by well-defined sets. Let $P$ be an approximation of the normal set $\Omega$, and we require $P\supset\Omega$ without loss of optimality. If we have some ${\bf z}\in P$ but ${\bf z}\notin \Omega$, it follows that ${\bf z}'\notin\Omega$ for any ${\bf z}'\succeq {\bf z}$ by the definition of a normal set. This implies that we can construct a better approximation $P'$ by cutting off a subset from $P$, i.e., $P' = P \setminus \{{\bf z}'\in P: {\bf z}'\succeq {\bf z}\}$. This ensures that $P\supset P' \supset \Omega$. Hence, the structural property of a normal set allows us to improve the approximation successively.

\begin{proposition}\label{pro_normal}
The feasible set $\Omega$ defined in \eqref{set_omega1} is normal.
\end{proposition}
The proof of Proposition~\ref{pro_normal} follows exactly the definition of a normal set and is similar to that in our previous work~\cite{Globecom16:Gong}. A simple way to approximate the normal set $\Omega$ is by generating regularly shaped \emph{polyblocks}, which is defined as follows:
\begin{definition}[\emph{Polyblock}]\label{def_polyblock}
A polyblock $P$ is a union of finite box sets, i.e., ${P}=\bigcup_{{\bf v}\in V}[{\bf 0},{\bf v}]$, where ${\bf v}$ is the vertex of $P$.
\end{definition}
It is clear that the optimum of an increasing function over a polyblock will be obtained on one of the vertex points. As the number of box sets increases, the polyblock will be a close approximation of~$\Omega$. Hence, we have the following proposition.
\begin{proposition}\label{pro_monot}
The optimum of throughput maximization problem in~\eqref{prob_simple_r} is attained on the upper boundary $\bar{\Omega}$.
\end{proposition}
Proposition~\ref{pro_monot} is rather intuitive and a formal proof easily follows from the results in~\cite{tuy}. The basic idea of the monotonic optimization algorithm is to approximate the feasible set $\Omega$ by iteratively generating regularly-shaped polyblocks. In one iteration of this process, e.g., the $k$-th iteration, the algorithm determines an upper bound $r_k^{\mathrm{U}}$ of the objective on one vertex of the polyblock $P_k$. It also updates a lower bound $r_k^{\mathrm{L}}$ by evaluating the objective $r(t,\gamma) = t\log(1+\gamma)$ on one upper boundary point $(t,\gamma)\in \bar\Omega$. Then, by trimming the polyblock $P_k$ and generating ``smaller" polyblock $P_{k+1}$,\footnote{Without causing confusions and for ease of presentation, we call that $P_{k+1}$ is ``smaller" than $P_k$ if $P_{k+1}\subset P_k$.} the iterative algorithm ensures that the gap between the upper and lower bounds is within the error distance $\epsilon$ after finite number of iterations. When the monotonic optimization algorithm converges to the optimal solution $(t,\gamma)$, we can retrieve the optimal solution $(t, {\bf c}, {\bf w}_{\mathrm{e}})$ to the problem in~\eqref{equ:original-form1}.

\subsection{Evaluate Upper and Lower Bounds on Problem~\eqref{equ:original-form1}}

Given the polyblock approximation $P_k$ in the $k$-th iteration, an upper bound $r^{\mathrm{U}}_k$ on the global optimum $r^* \triangleq \max_{{\bf z}\in \Omega}r({\bf z})$ is given by $r^{\mathrm{U}}_k=\max_{{\bf z}\in P_k}\,\,r({\bf z})$. Let
\[
{\bf z}_k  \triangleq (t_k,\gamma_k) =\arg\max_{{\bf v}\in V_k}r({\bf v}),
\]
where $V_k$ denotes the vertex set of $P_k$. Then, $r_k^{\mathrm{U}}=r({\bf z}_k)$ is the upper bound on $r^*$. If ${\bf z}_k\in \Omega$, the upper bound $r({\bf z}_k)$ also serves a lower bound and thus the algorithm terminates with the objective $r({\bf z}_k)$. In the case of ${\bf z}_k\notin\Omega$, a feasible lower bound on $r^*$ can be found by projection, i.e., we scale the vertex ${\bf z}_k$ by a factor $\lambda\in(0,1)$ such that the projection point, denoted by ${\bf o}_k(\lambda) \triangleq \lambda {\bf z}_k$, is on the upper boundary $\bar{\Omega}$. This suggests a bisection method to find the optimal scaling factor $\lambda_k = \arg\max\{\lambda: {\bf o}_k(\lambda)\in\bar\Omega\}$, i.e., we first check the feasibility ${\bf o}_k(\lambda) \in\Omega$ with a fixed $\lambda$, then increase $\lambda$ if ${\bf o}_k(\lambda) \in\Omega$ or decrease it otherwise. The feasibility check is essential for the convergence of the bisection method. It is problem-dependent and our main challenge in solving the throughput optimization problem.

Given the scaling factor $\lambda$ and vertex ${\bf z}_k = (t_k,\gamma_k)$ in the $k$-th iteration, checking feasibility of ${\bf o}_k(\lambda) \in \Omega$ is equivalent to finding a feasible solution $({\bf c},{\bf W}_{\mathrm{e}})$ such that
\begin{subequations}\label{equ_fprj}
\begin{align}
	{\bf c}^T({\bf A}-\lambda \gamma_k {\bf B}){\bf c} &\,\geq \, \lambda \gamma_k, \label{equ_prj1}\\
\max_{\mathbb{P} \in \mathcal{P}_m}\mathbb{P}\left(\phi_m({\bf c}) \geq \bar{\phi}_m\right) &\,\leq\, \zeta, \quad \forall m\in\mathcal{C}, \label{con_ip_exp}\\
(1 - 2 \lambda t_k) \eta p_{\mathrm{o}} {\bf f}^H_n {\bf W}_{\mathrm{e}} {\bf f}_n & \geq p_n \lambda t_k, \quad \forall n \in \mathcal{N}. \label{equ_prj2}
\end{align}
\end{subequations}
To proceed, we can firstly evaluate the maximum of a quadratic optimization as follows:
\begin{align}\label{pro_projection}
q(\lambda,{\bf z}_k) \triangleq \max_{{\bf c},{\bf W}_{\mathrm{e}}} \{ {\bf c}^T({\bf A}-\lambda \gamma_k {\bf B}){\bf c} : \eqref{con_ip_exp}-\eqref{equ_prj2}\},
\end{align}
and then compare the maximum $q(\lambda,{\bf z}_k)$ to the target $\lambda \gamma_k$. If $q(\lambda,{\bf z}_k)\geq \lambda \gamma_k$, we have $ \lambda {\bf z}_k\in\Omega$ and then increase the scaling factor $\lambda$ in the next iteration. The difficulty of problem (\ref{pro_projection}) lies in the indefinite matrix coefficient $({\bf A}-\lambda \gamma_k {\bf B})$ in the objective and the non-convex probabilistic constraint in \eqref{con_ip_exp}.
\begin{proposition}\label{pro_equvi_projection}
The maximum objective $q(\lambda ,{\bf z}_k)$ in (\ref{pro_projection}) can be evaluated by the following equivalence\footnote{For notational convenience and conciseness, the auxiliary or intermediate variables are not listed as the design variables of the optimization problem. The same convention applies to the following optimization problems, e.g., \eqref{equ:orignal-form} and \eqref{equ:final-form}.}:
\begin{subequations}\label{prob_prjcvx}
\begin{align}
\max_{{\bf c},{\bf W}_{\mathrm{e}}}~&~ {\bf c}^T({\bf A}-\lambda \gamma_k {\bf B}){\bf c} \label{obj_prjcvx} \\
\text{s.t.} ~&~ \lambda t_k c_n^2 \leq (1 - 2 \lambda t_k) \eta p_{\mathrm{o}} {\bf f}^H_n {\bf W}_{\mathrm{e}} {\bf f}_n, \quad \forall n\in\mathcal{N}, \label{con3_prjcvx}\\
~&~{\bf M }_m\succeq \left[\begin{array}{cc} \mathrm{D}({\bf c}\circ{\bf c}) & 0 \\ 0 & \nu_m - \bar{\phi}_m \end{array}\right], \quad \forall m\in\mathcal{C}, \label{con2_prjcvx}\\
~&~ {\bf Tr}({\bf \Sigma}_m{\bf M}_m) \leq \nu_m \zeta,\,\, {\bf M}_m \succeq 0, \quad \forall m\in\mathcal{C}, \label{con1_prjcvx}\\
~&~ {\bf c}\succeq {\bf 0} \text{ and } {\bf W}_{\mathrm{e}}\succeq {\bf 0}.
\end{align}
\end{subequations}
\end{proposition}
The equivalence relies on the convex reformulation of chance constraints in~\eqref{con_ip_exp}. The detailed proof of Proposition~\ref{pro_equvi_projection} is relegated to Appendix~\ref{proof_equvi_projection}. We can further introduce a rank-one matrix ${\bf C}={\bf c}{\bf c}^T$ to linearize the objective function in \eqref{obj_prjcvx}. Note that ${\bf c}^T({\bf A}-\lambda \gamma_k {\bf B}){\bf c} = \textbf{Tr}({\bf C}({\bf A}-s\gamma_k {\bf B}))$ and $\text{D}({\bf c}\circ{\bf c})={\bf \Delta}({\bf C})$, where ${\bf \Delta}({\bf C})$ is the diagonal matrix by setting all off-diagonal elements to zeros. By this convention, we can transform~\eqref{con2_prjcvx} into a linear matrix inequality. As an approximation, we drop the rank-one constraint on ${\bf C}$ and obtain the SDR representation of (\ref{prob_prjcvx}), which now can be solved efficiently by the interior-point algorithms. If the optimal ${\bf C}^\star$ happens to be rank one, a feasible solution ${\bf c}^\star$ to~(\ref{pro_projection}) can be extracted by eigen-decomposition. Otherwise, we can extract an approximate rank-one solution by Gaussian randomization~\cite{rankluo}. What worth mentioning is that the nonlinear EH model developed in~\cite{noneh} can be also applied to the power budget constraint in~\eqref{con3_prjcvx} without changing the problem structure. Let $E_n({\bf w}_e)$ denote the energy harvested by relay-$n$ from the HAP. Its nonlinear dependency on ${\bf w}_e$ can be modeled as a logistic function. That is, $E_n({\bf w}_e) = (1-\Omega_n)^{-1} (\ell_n({\bf w}_e) -\nu_n  \Omega_n)$ where $\ell_n({\bf w}_e) \triangleq \nu_n \left( 1 + e^{-a_n \left( p_{\mathrm{o}} |{\bf f}_n^H {\bf w}_e|^2 - b_n \right)} \right)^{-1}$ is the logistic function representing the nonlinearity in EH. The constants $\nu_n$ and $(a_n, b_n)$ can be estimated via measurement data. The constant $\Omega_n$ ensures that $E_n({\bf w}_e)=0$ when $p_{\mathrm{o}} = 0$. With this nonlinear EH model, we can replace $\eta p_{\mathrm{o}} |{\bf f}^H_n {\bf w}_{\mathrm{e}}|^2$ in~\eqref{con3_prjcvx} by the nonlinear function $E_n({\bf w}_e)$ and thus revise the power budget constraint as follows:
\begin{equation}\label{equ_non_bgt}
\lambda t_k c_n^2 \leq (1 - 2 \lambda t_k)  E_n({\bf w}_e) , \quad \forall n\in\mathcal{N}.
\end{equation}
We can verify that the above substitution will not alter the solution method to problem~\eqref{prob_prjcvx}. Specifically, after substitution and some manipulations, we can reformulate~\eqref{equ_non_bgt} as follows:
\begin{equation}\label{eq-nonlinear-ts}
p_{\mathrm{o}} |{\bf f}_n^H {\bf w}_e|^2 \geq b_n + a_n^{-1} \log \left( \frac{(1-\Omega_n) p_n' + \nu_n \Omega_n}{(1-\Omega_n)(\nu_n - p_n')} \right),
\end{equation}
where $p_n' = \frac{\lambda t_k}{1 - 2\lambda t_k} c^2_n$. Till this point, we can easily verify that \eqref{eq-nonlinear-ts} defines a convex feasible set and therefore the revised maximization problem in~\eqref{prob_prjcvx} can still be solved efficiently by the interior-point algorithms.

\subsection{Update a Smaller Polyblock}
The projection point ${\bf o}_k(\lambda)$ allows us to trim the polyblock $P_k$ and generate a ``smaller" polyblock. Specifically, when ${\bf z}_k \neq {\bf o}_k$, we define $\Delta_k  =\{{\bf z}\in P_k \,|\,{\bf z}\succeq{\bf o}_k\}$, and then we can generate a new polyblock $P_{k+1}$ by cutting off $\Delta_k$ from the polyblock $P_k$, i.e., $P_{k+1}\triangleq P_k \setminus \Delta_k$. It is easy to verify that the new polyblock $P_{k+1}$ will lead to a tighter upper bound on $r^*$. That is, $P_{k}\supset P_{k+1}\supset \Omega$. Moreover, the cutting off $\Delta_k$ from $P_k$ will erase some vertices in $V_k$ but also generate new ones. Let $\overline{V}_k$ denote the set of vertices that are erased from $V_k$. Then, the vertex set $V_{k+1}$ can be updated by
\begin{equation*}\label{equ_vertex_set}
V_{k+1}=(V_k \setminus \overline{V}_k) \cup V^{k+1},
\end{equation*}
where $V^{k+1}$ denotes the set of newly generated vertices. A discussion on the new vertex set $V_{k+1}$ is detailed in our previous work~\cite{Globecom16:Gong}. Till this point, the new polyblock can be constructed easily by $P_{k+1}=\bigcup_{{\bf v} \in V_{k+1}} [{\bf 0}, {\bf v}]$.

\begin{algorithm}[t]
\caption{Joint Optimization of Beamforming and Relaying Strategies with the TS Scheme}\label{alg_general}
\begin{algorithmic}[1]

\State \hspace{-0.1cm} Initialize vertex point ${\bf v}_0= (1/2, {\bf B}^{-1}{\bf A})$

\State \hspace{-0.1cm} Set $P_0 = [{\bf 0}, {\bf v}_0]$, $V_0 = \{{\bf v}_0\}$, and $\epsilon = 10^{-5}$

\State \hspace{-0.1cm} Set $r_k^{\mathrm{U}}=1$ and $r_k^{\mathrm{L}}=0$ for $k=0$

\State \hspace{-0.1cm} {\bf while} $| r_k^{\mathrm{U}} - r_k^{\mathrm{L}} |\geq \epsilon$

\State \hspace{0.4cm} $k\leftarrow k+1$, ${\bf z}_k\leftarrow\arg\max_{{\bf v}\in V_{k-1}} r({\bf v})$, $r_k^{\mathrm{U}}\leftarrow r({\bf {\bf z}_k})$

\State \hspace{0.4cm} {\bf if} ${\bf z}_k\in\Omega$ {\bf then} $r_k^{\mathrm{L}} \leftarrow r({\bf z}_k)$, ${\bf z}^*\leftarrow{\bf z}_k$

\State \hspace{0.4cm} {\bf else}


\State \hspace{0.9cm} $\lambda_{\min} \leftarrow 0$, $\lambda_{\max} \leftarrow 1$, $\lambda \leftarrow (\lambda_{\min} + \lambda_{\max})/2$

\State \hspace{0.9cm} {\bf while} $|\lambda_{\max} -  \lambda_{\min}|\geq \epsilon$

\State \hspace{1.3cm} Update $q(\lambda,{\bf z}_k)$ by the optimum to SDP \eqref{prob_prjcvx}

\State \hspace{1.3cm} {\bf if} $q(\lambda,{\bf z}_k)\geq \lambda \gamma_k$, {\bf then}  $ \lambda_{\min} \leftarrow \lambda$ {\bf else} $ \lambda_{\max} \leftarrow \lambda$  {\bf end if}

\State \hspace{0.9cm} {\bf end while}


\State \hspace{0.9cm} {\bf if} $r({\bf o}_k)\geq r_k^{\mathrm{L}}$ {\bf then} $r_k^{\mathrm{L}} \leftarrow r({\bf o}_k)$, ${\bf z}^*\leftarrow{\bf o}_k$ {\bf end if}

\State \hspace{0.9cm} Update vertex set: $V_{k}\leftarrow (V_{k-1} \setminus \overline{V}_{k-1}) \cup V^{k}$

\State \hspace{0.9cm} Construct new polyblock: $P_{k}\leftarrow\bigcup_{{\bf v} \in V_{k}} [{\bf 0}, {\bf v}]$

\State \hspace{0.4cm} {\bf end if}

\State \hspace{-0.1cm} {\bf end while}

\end{algorithmic}

\end{algorithm}

We present the customized monotonic optimization algorithm in Algorithm~\ref{alg_general} to solve the throughput optimization~\eqref{equ:original-form1} with the TS scheme. The stop criterion $\epsilon$ is an error tolerance that ensures that the algorithm returns a close to optimal solution when it terminates. The initial polyblock $P_0$ can be large enough by evaluating the maximum of $t$ and $\gamma$. It is obvious that $t\leq 1/2$ and $\gamma  \leq {\bf B}^{-1}{\bf A}$. Hence, we can set the initial vertext point as ${\bf v}_0 = (1/2, {\bf B}^{-1}{\bf A})$. Algorithm~\ref{alg_general} can be implemented at the HAP after information exchange with the relays and the receiver. The information required for the algorithm is the matrix coefficients ${\bf A}$ and ${\bf B}$, relating to the relays' channel conditions and EH capabilities, which can be acquired through information exchange at the beginning of each time slot. In particular, the HAP can broadcast a known pilot to all relays, who can estimate the channel coefficients locally and simultaneously. Thus, we can reduce the signaling overhead by such parallelized channel estimations at individual relays. After channel estimation, the relays can feedback the channel information to the HAP in a time-slotted manner. Meanwhile, the receiver can sequentially estimate the channel from each relay to the receiver.

\begin{figure}[t]
 \centering
 \includegraphics[width=\singlesize\textwidth]{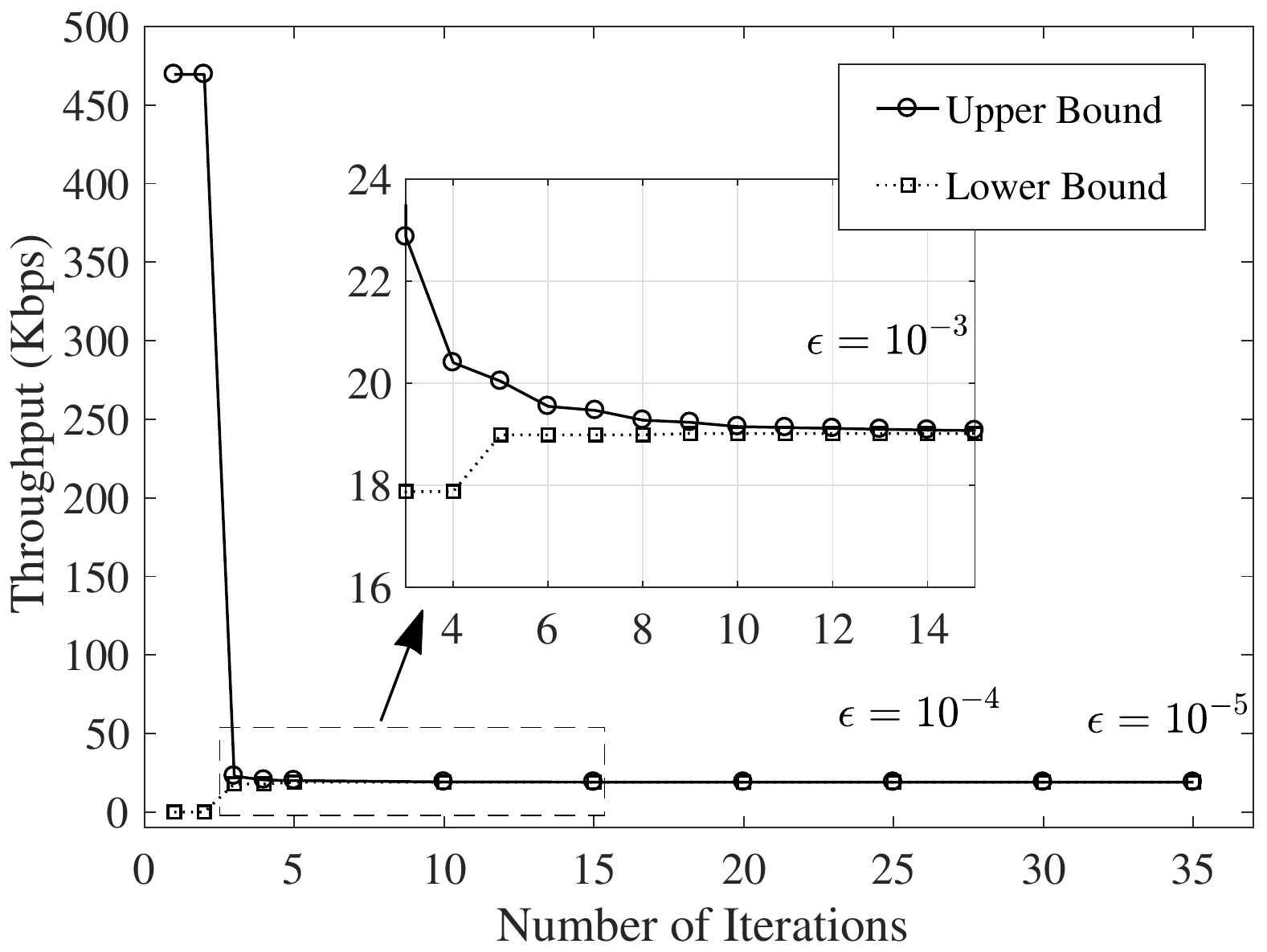}\\
 \caption{{Convergence of the upper and lower bounds in Algorithm~\ref{alg_general}.}}\label{fig:alg-conv}
\end{figure}

Fig.~\ref{fig:alg-conv} gives a numerical illustration of the convergence of Algorithm~\ref{alg_general}. We set $p_{\mathrm{o}}=50$ mW and $\zeta=0.3$. The upper and lower bounds converge quickly within few iterations. The gap between the upper and lower bounds decreases to $\epsilon=10^{-3}$, $10^{-4}$, $10^{-5}$ within 13, 25, and 35 iterations, respectively. In particular, the gap decreases quickly to $10^{-3}$ and the throughput has no significant improvement afterwards. This implies that good performance can be obtained from Algorithm~\ref{alg_general}, by setting a relatively larger $\epsilon$ (e.g., $\epsilon=10^{-3}$) to ensure fast convergence. The computational complexity of Algorithm~\ref{alg_general} mainly lies in two parts. One part relates to the outer-loop iterations of monotonic optimization and the other part lies in the solution to SDP problems. The complexity of monotonic optimization is generally increasing exponentially in the dimension of decision variable ${\bf z}$. However, by a change of variables and a reformulation of~\eqref{equ:original-form1}, we ensure that the dimension of ${\bf z}$ is fixed at~2. By minimizing the search space, Algorithm~\ref{alg_general} improves the efficiency of monotonic optimization algorithm even with a larger number of relays and CUEs. It is observed from Fig.~\ref{fig:alg-conv} that the outer-loop requires few iterations to achieve a desired accuracy. Therefore, the overall complexity of Algorithm~\ref{alg_general} depends on the solution to SDPs. By the analytical work in~\cite{inter_point}, the computational complexity of SDPs can be evaluated by ${\cal O}(N^{1.5}M^4 + N^{6.5})$, where $N$ and $M$ denote the number of relays and the number of the HAP's antennas, respectively.

\section{Robust Throughput Maximization with the PS Scheme}\label{sec_ps}
\label{sec:optimization_ps}

Alternative to the TS scheme, the HAP and relays can adopt the PS scheme which leads to a different problem formulation. In the sequel, we show that the throughput maximization with the PS scheme can be approximated by a monotonic optimization problem and solved similarly as that of problem~\eqref{equ:original-form1}.

With the PS scheme, each relay $n\in\mathcal{N}$ can adjust its PS ratio, denoted by $\rho_n$. The vector of all relays' PS ratios is given by $\boldsymbol{\rho} = [\rho_1,\ldots,\rho_n,\ldots,\rho_N]^T$. Let ${\bf w}_{\mathrm{p}}$ denote the normalized beamforming vector of the HAP for mixed information and power transfer in the first hop. Given the beamforming signal ${\bf x}_s = \sqrt{p_{\mathrm{o}}}{\bf w}_{\mathrm{p}}s$, the received information bearing signal at the relay-$n$ is given by $y_n = \sqrt{(1-\rho_n) p_{\mathrm{o}}} \mathbf{f}_n^H {\bf w}_{\mathrm{p}} s$, and its power amplifying coefficient can be rewritten as
\begin{equation}\label{equ_ampps}
	x_n = \left(\frac{p_n}{(1-\rho_n) p_{\mathrm{o}} \mathbf{f}_n^H\mathbf{W}_{\mathrm{p}}\mathbf{f}_n + 1}\right)^{1/2},
\end{equation}
where ${\bf W}_{\mathrm{p}} = {\bf w}_{\mathrm{p}}{\bf w}_{\mathrm{p}}^H$ denotes transmit covariance of the the HAP and $p_n$ is normalized transmit power of relay-$n$. Again, our objective is to maximize the throughput in a time slot $r = \log (1 + \gamma(\boldsymbol{\rho},{\bf p},{\bf W}_{\mathrm{p}}))$ by optimizing the HAP's beamforming strategy $\mathbf{W}_{\mathrm{p}}$, the relays' PS ratio $\boldsymbol{\rho}$ and transmit power ${\bf p}$:
\begin{subequations}\label{equ:orignal-form}
\begin{align}
\max_{\boldsymbol{\rho}, {\bf p}, {\bf W}_{\mathrm{p}}}~~ & \log\left(1 + \frac{ |({\bf x}\circ{\bf y})^H{\bf g}|^2}{1+ {\bf x}^T \text{D}({\bf g}\circ{\bf g}){\bf x}}\right) \label{equ:obj}\\
\text{s.t.}~~ & \max_{\mathbb{P}\in \mathcal{P}_m} \mathbb{P} \left(\phi_m({\bf p}) \geq \bar{\phi}_m \right) \leq \zeta, \quad \forall m\in\mathcal{C}, \label{con_interps}\\
~~& p_n \leq \eta p_{\mathrm{o}} \rho_n\mathbf{f}_n^H {\bf W}_{\mathrm{p}}\mathbf{f}_n, \quad \forall n \in \mathcal{N}, \label{con_powps}\\
~~& 0 \preceq {\boldsymbol{\rho}} \preceq 1, {\bf p}\succeq {\bf 0}, \text{ and } \textbf{Tr}({\bf W}_{\mathrm{p}})\leq 1. \label{con_rank}
\end{align}
\end{subequations}
where \eqref{con_interps} and \eqref{con_powps} define the interference constraints at CUEs and the power budget constraints at the relays, respectively. Note that the channel time for information transmission in the PS scheme is constant and thus omitted in~\eqref{equ:obj}.

\subsection{Approximation and Monotonicity}

The power splitting makes the coupling between SNR $\gamma$ and the HAP's beamforming strategy ${\bf W}_{\mathrm{p}}$ more complicated. We are unable to transform the problem in \eqref{equ:orignal-form} into an equivalent monotonic optimization problem. To this end, we consider a relaxation to simplify the objective function in \eqref{equ:obj}.
\begin{proposition}\label{pro:reform}
Define $X = {\bf x}^T\text{D}({\bf g \circ \bf g}){\bf x}$ and $Y = {\bf y}^T{\bf y}$. Let $\theta$ denote a non-zero scalar. The problem in (\ref{equ:orignal-form}) is lower bounded by the solution to the following problem.
\begin{subequations}\label{equ:reformulation}
\begin{align}
\max_{X,Y} ~&~ XY(1 + X)^{-1} \label{obj_xy}\\
\text{s.t.}~&~ \mathbf{y} = \theta \mathbf{x} \circ \mathbf{g}, \text{ and } \eqref{con_interps}-\eqref{con_rank}		.	 \label{con_xy}
\end{align}
\end{subequations}
\end{proposition}
The proof of Proposition~\ref{pro:reform} is straightforward by applying \emph{Cauchy inequality}: $\gamma(\mathbf{x}, \mathbf{y}) \leq \frac{\left[{\bf x}^T\text{D}({\bf g}\circ{\bf g}){\bf x}\right] {\bf y}^T{\bf y} }{1+ {\bf x}^T\text{D}({\bf g}\circ{\bf g}){\bf x} } = XY(1+X)^{-1}$. The equality holds if and only if $\mathbf{y}= \theta \mathbf{x} \circ \mathbf{g}$ for some $\theta \neq 0$. By imposing the equality condition to the original problem, it is equivalent to maximize $XY(1+X)^{-1}$ in~\eqref{obj_xy}. It is clear that~\eqref{equ:reformulation} imposes more stringent requirements on the feasible set of $(\boldsymbol{\rho},{\bf p},{\bf W}_{\mathrm{p}})$. For any solution feasible to \eqref{equ:reformulation}, it is also feasible to~\eqref{equ:orignal-form} and thus gives a lower bound on \eqref{equ:orignal-form}. The new objective $\gamma(X,Y)$ is increasing in both $X$ and $Y$,\footnote{We use $\gamma(X,Y)$ and $\gamma(\boldsymbol{\rho},{\bf p}, {\bf W}_{\mathrm{p}})$ interchangeably to denote the SNR performance with the PS scheme.} implying that the optimum will be achieved on the boundary of the feasible region, which is defined as follows:
\begin{equation}\label{equ:feasible-region}
\Omega_{\mathrm{p}} = \left\{ (X,Y) \left|
\begin{aligned}
&X \leq {\bf x}^T\text{D}({\bf g}\circ{\bf g}){\bf x} \text{ and } Y \leq {\bf y}^T{\bf y} \\
&\quad \forall ({\bf x},{\bf y}) \text{ feasible to } \eqref{con_xy}\end{aligned}
\right\}\right..
\end{equation}
To proceed with the monotonic optimization algorithm, we further verify that $\Omega_{\mathrm{p}}$ is also a normal set.
\begin{proposition}\label{pro:normal}
The feasible set $\Omega_{\mathrm{p}}$ defined in \eqref{equ:feasible-region} is normal.
\end{proposition}

The proof of Proposition~\ref{pro:normal} is given in Appendix~\ref{apd:proof-normal}. In Fig.~\ref{fig:feasible-region-shri}, through a numerical example, we show the feasible region~$\Omega_{\mathrm{p}}$ and evaluate how the probability limit~$\zeta$ affects the shape of~$\Omega_{\mathrm{p}}$. We fix $p_{\mathrm{o}}=30$mW and vary $\zeta$ from $0.1$ to $0.3$. The channel model and network configuration are the same as those used in the simulation presented in Section~\ref{sec:simulation}. We observe that the size of $\Omega_{\mathrm{p}}$ shrinks when $\zeta$ decreases, which implies a more stringent interference requirement at the CUEs. Moreover, with different $\zeta$, the shape of $\Omega_{\mathrm{p}}$ always demonstrates a non-convex structure. To revise Algorithm~\ref{alg_general} applicable for the PS scheme, the initial polyblock can be set by finding the upper bounds on $X$ and $Y$, respectively. Let $\bar g_{\max} \triangleq \max_{n\in\mathcal{N}} |g_n|^2$ and $\bar f \triangleq \sum_{n\in \mathcal{N}}||{\bf f}_n||^2$. By the construction of $X ={\bf x}^T\text{D}({\bf g \circ \bf g}){\bf x}$ and the relations in \eqref{equ_ampps} and \eqref{con_powps}, we have $X  \leq \bar g_{\max} ||{\bf x}||^2 = \bar g_{\max} \sum_{n\in\mathcal{N}} \frac{p_n}{1+|y_n|^2} \leq \eta \bar g_{\max}\bar f p_{\mathrm{o}}$. Similarly, we have $Y \leq \bar f p_{\mathrm{o}}$. Hence, the initial vertex can be set as $( \eta \bar g_{\max} \bar f p_{\mathrm{o}} , \bar f p_{\mathrm{o}} )$.

\begin{figure}[t]
 \centering
 \includegraphics[width=\singlesize\textwidth]{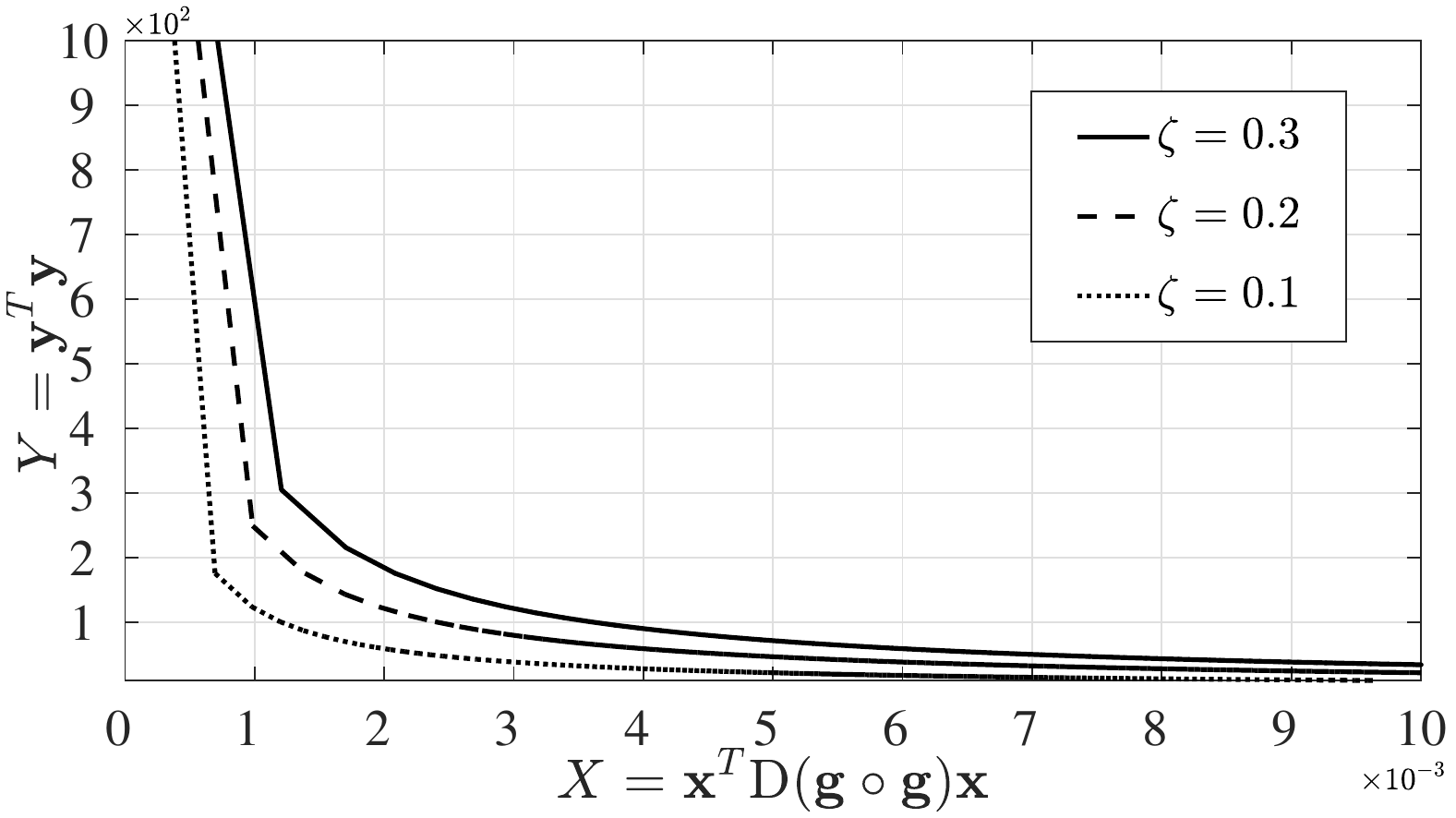}\\
 \caption{{The non-convex and normal feasible set $\Omega_{\mathrm{p}}$.}}\label{fig:feasible-region-shri}
\end{figure}

\subsection{Update Lower Bound on \eqref{equ:reformulation}}

In the $k$-th iteration of the monotonic optimization algorithm, assuming that ${\bf z}_k = (X_k, Y_k) \notin \Omega_{\mathrm{p}}$ is the upper-bounding vertex of the polyblock $P_k$, we need to find the maximum scaling factor $\lambda_k$ such that the scaled vertex $\lambda_k {\bf z}_k$ is feasible to $\Omega_{\mathrm{p}}$, i.e., $(\lambda_k X_k, \lambda_k Y_k) \in \bar\Omega_{\mathrm{p}}$. This is equivalent to find a solution $(\boldsymbol{\rho}, {\bf p}, {\bf W}_{\mathrm{p}})$ satisfying the following conditions:
\begin{subequations}\label{equ:feasible-form}
\begin{align}
||{\bf y}||^2      ~ & \geq~ \lambda_k Y_k, \label{equ:B-bound}\\
{\bf x}^T\text{D}({\bf g}\circ{\bf g}){\bf x}  ~ & \geq ~ \lambda_k X_k ,\label{equ:A-bound}\\
\theta_k\mathbf{x}\circ \mathbf{g}
~&= ~{\bf y}, \label{equ:equ-constr}\\
\eta \rho_n p_{\mathrm{o}} \mathbf{f}_n^H{\bf W}_{\mathrm{p}}\mathbf{f}_n ~& \geq ~ p_n, \quad \forall n \in \mathcal{N}, \label{equ:power-constr}\\
\max_{\mathbb{P}\in \mathcal{P}_m} \mathbb{P} \left(\phi_m({\bf p}) \geq \bar{\phi}_m \right)~& \leq ~ \zeta, \quad \forall m \in \mathcal{C}.\label{equ:inter-constr-form}
\end{align}
\end{subequations}
Additionally, we can show that $\theta_k$ can be set as follows.
\begin{proposition}\label{pro:optimal-C}
Given ${\bf z}_k = (X_k, Y_k)\succeq {\bf 0}$ and $X_k\neq 0$, the coefficient $\theta_k$ in \eqref{equ:feasible-form} can be set to $\theta_k=\sqrt{Y_k/X_k}$.
\end{proposition}
The proof of proposition~\ref{pro:optimal-C} is relegated to Appendix~\ref{apd:proof-optimal-C}. Substituting (\ref{equ:equ-constr}) into (\ref{equ:B-bound}), the inequality in (\ref{equ:B-bound}) is exactly the same as \eqref{equ:A-bound}. Hence, the feasibility of \eqref{equ:feasible-form} can be evaluated by maximizing $||{\bf y}||^2 $ as follows:
\begin{equation}\label{equ:feasible-reform-1}
\tilde q(\mathbf{z}_k) \triangleq \max_{\boldsymbol{\rho},{\bf p}, {\bf W}_{\mathrm{p}}}\{ ||{\bf y}||^2 : \eqref{equ:equ-constr}-\eqref{equ:inter-constr-form}\}.
\end{equation}
Once~\eqref{equ:feasible-reform-1} is solved, we can simply set the scaling factor by $\lambda_k = \tilde q(\mathbf{z}_k)/Y_k$. Otherwise, we have $\lambda Y_k > \tilde q(\mathbf{z}_k)$, and thus $\lambda {\bf z}_k\notin \Omega_{\mathrm{p}}$ for any $\lambda>\lambda_k$. Though it is still difficult to solve (\ref{equ:feasible-reform-1}) due to its non-convex structure, we can present a convex reformulation as follows by introducing auxiliary variables:
\begin{subequations}\label{equ:final-form}
\begin{align}
\max_{ {\bf p}, {\bf W}_{\mathrm{p}},\bar{\bf W}_{\mathrm{p}}} ~& || \boldsymbol{\kappa}||_1 \\
\text{s.t.}
~& p_n \leq \eta p_{\mathrm{o}} \mathbf{f}_n^H \bar{\bf W}_{\mathrm{p}} \mathbf{f}_n,\quad \forall\,n\in\mathcal{N}, \\
~& \left[
      \begin{matrix}
      p_n \theta_k^2 g_n^2 - \kappa_n & \kappa_n\\
      \kappa_n         & 1
      \end{matrix}
      \right] \succeq 0,\quad \forall\,n\in\mathcal{N}, \label{con_lmi}\\
~& \kappa_n \leq p_{\mathrm{o}} \mathbf{f}_n^H (\mathbf{W}_{\mathrm{p}} - \bar{\mathbf{W}}_{\mathrm{p}}) \mathbf{f}_n,\quad \forall\,n\in\mathcal{N},\\
~& {\bf M}_m \succeq \left[
      \begin{matrix}
      \text{D}( {\bf p}) & 0\\
      0         & v_m-\bar{\phi}_m
      \end{matrix}
      \right],\quad \forall\,m\in\mathcal{C}, \label{con_prob1}\\
~& {\bf Tr}(\mathbf{\Sigma}_m{\bf M}_m) \leq v_m \zeta, \quad \forall\,m\in\mathcal{C}, \label{con_prob2} \\
~&  {\bf M}_m \succeq 0, \text{ and } v_m\geq 0, \quad \forall\,m\in\mathcal{C},\label{con_prob3}\\
~& {\bf Tr}({\bf W}_{\mathrm{p}}) \leq 1, \text{ and } {\bf Tr}(\bar{\bf W}_{\mathrm{p}}) \leq 1,\\
~& {\bf p}\succeq {\bf 0},   {\bf W}_{\mathrm{p}}\succeq{\bf 0}, \bar{\bf W}_{\mathrm{p}}\succeq{\bf 0} , \text{ and } \boldsymbol{\kappa} \succeq {\bf 0}.
\end{align}
\end{subequations}
The equivalence between~\eqref{equ:feasible-reform-1} and~\eqref{equ:final-form} relies on the convex reformulations of the constraints in~\eqref{equ:equ-constr}-\eqref{equ:inter-constr-form}. If ${\bf W}_{\mathrm{p}}^{*}$ and $\bar{\bf W}_{\mathrm{p}}^{*}$ are optimal to \eqref{equ:final-form}, we can simply set the optimal PS ratio as $\rho_n^* = \mathbf{f}_n^H \bar{\mathbf{W}}_{\mathrm{p}}^{*} \mathbf{f}_n/ \mathbf{f}_n^H {\mathbf{W}_{\mathrm{p}}^{*}} \mathbf{f}_n$ for any $n \in \mathcal{N}$. The detailed analysis is presented in Appendix~\ref{proof_equvi_ps}. Given the solution to~\eqref{equ:final-form}, we can determine the maximum scaling factor $\lambda_k$ and correspondingly the projection point $\lambda_k{\bf z}_k$ in the $k$-th iteration. Then, we can construct a smaller polyblock~$P_{k+1}$ following similar procedures as that in Algorithm~\ref{alg_general}. The revised procedures for the PS scheme are listed in Algorithm~\ref{alg:third}.

\begin{algorithm}[!hbt]
\caption{Joint Optimization of Beamforming and Relaying Strategies with the PS Scheme}\label{alg:third}
\begin{algorithmic}[1]

\State \hspace{-0.1cm} Collect channel information $({\bf f}_n, {\bf z}_n)$ and ${\bf g}$

\State \hspace{-0.1cm} Initialize vertex point ${\bf v}_0= ( \eta \bar g_{\max} \bar f p_{\mathrm{o}} , \bar f p_{\mathrm{o}} )$

\State \hspace{-0.1cm} Set $P_0 = [{\bf 0}, {\bf v}_0]$, $V_0 = \{{\bf v}_0\}$, and $\epsilon = 10^{-5}$

\State Set $\gamma_k^{\mathrm{U}}= p_{\mathrm{o}} \sum_{n=1}^{N}\mathbf{f}_n^H \mathbf{f}_n$ and $\gamma_k^{\mathrm{L}}=0$ for $k=0$
\While{$|\gamma_k^{\mathrm{U}} - \gamma_k^{\mathrm{L}}| \geq \epsilon$}
\State $k \leftarrow k + 1$, $\mathbf{z}_k \leftarrow \arg \max_{\mathbf{v}\in V_{k-1}} \gamma(\mathbf{v})$, $\gamma_k^{U}\leftarrow\gamma(\mathbf{z}_k)$
\State $\theta_k \leftarrow \sqrt{Y_k/X_k}$
\If {$\mathbf{z}_k \in \Omega$} $\gamma_k^{\mathrm{L}} \leftarrow \gamma(\mathbf{z}_k)$, $\mathbf{z}^* \leftarrow \mathbf{z}_k$
\Else
\State Find projection $\mathbf{o}_k = \lambda_k \mathbf{z}_k$ by solving SDP (\ref{equ:final-form})
\State {\bf if} $\gamma(\mathbf{o}_k) \geq \gamma_k^{\mathrm{L}}$ {\bf then} $\gamma_k^{\mathrm{L}} \leftarrow \gamma(\mathbf{o}_k)$, $\mathbf{z}^* \leftarrow \mathbf{o}_k$ {\bf end if}
\State Update vertex set $V_k$ and polyblock $P_k$
\EndIf
\EndWhile
\State Retrieve ($\boldsymbol{\rho}^*$, $\mathbf{p}^*$, $\mathbf{W}^*$) from the solution to (\ref{equ:final-form})
\end{algorithmic}
\end{algorithm}

\section{Performance Evaluation}\label{sec:simulation}
We firstly compare the TS and PS schemes in terms of throughput performance and energy efficiency. The results reveal that the PS scheme performs generally better than the TS scheme. Then, focusing on the PS scheme, we investigate how different control parameters have impacts on different performance measures. In particular, we show the joint control of the HAP's beamforming, the relays' PS ratios and transmit power to improve energy efficiency. We further vary the range of channel uncertainty and examine its effects on the joint control at the relays. Specifically, without loss of generality, we set $K = 3$ antennas at the HAP and $N = 3$ relays for each DUE. We only consider one CUE in the simulations, but the results can be easily extended to the case with multiple CUEs, ensuring the protection for the most vulnerable CUE. The noise power density is set to -90 dBm and the bandwidth is 100 kHz. The energy conversion efficiency is set to $\eta = 0.5$. The HAP's transmit power $p_{\mathrm{o}}$ in milliwatts is limited by 50 mW. The pass loss $L$ is modeled by the log-distance propagation model: $L=L_0+10\alpha\log_{10}(d/d_0)$, where we set the path loss exponent as $\alpha=2$ and the path loss at unit distance $d_0=1$ as $L_0=25$ dB. To differentiate different relays, we set the distances in meters from HAP to the relays as ${\bf d}_{\bf f}=[2, 3, 4]$, which implies that relay-1 generally has the most preferable channel condition. For a fair comparison, we set the same distances from individual relays to the DRx and the CUE, given by ${\bf d}_{\bf g}=[2,2,2]^T$ and ${\bf d}_{\bf z}=[3,3,3]^T$, respectively.

\begin{figure}[t]
 \centering
 \subfloat[Dynamics of throughput performance]{\includegraphics[width=\singlesize\textwidth]{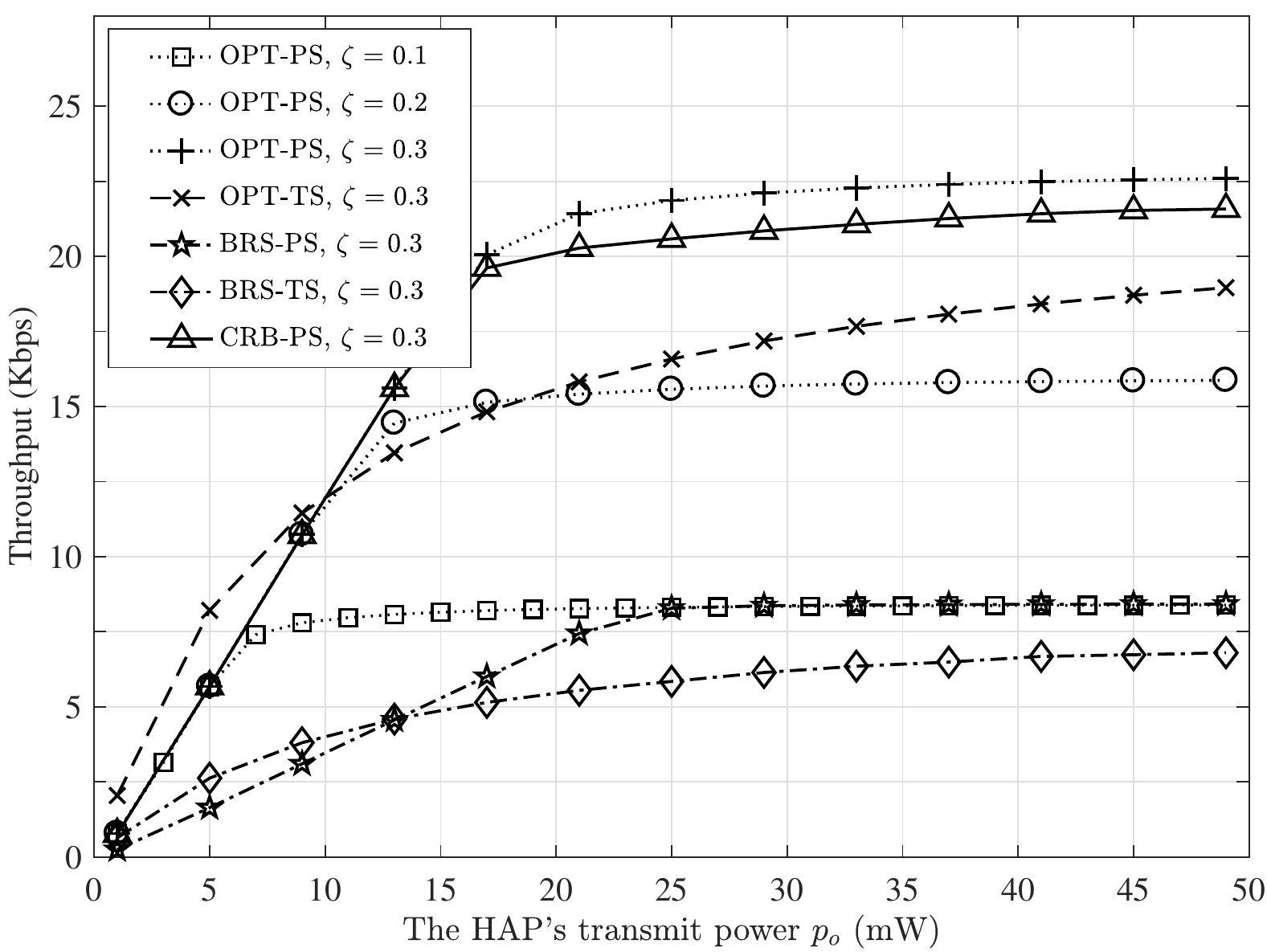}\label{fig:comp-throughput}}\\
 \subfloat[The PS scheme is generally more energy efficient]{\includegraphics[width=\singlesize\textwidth]{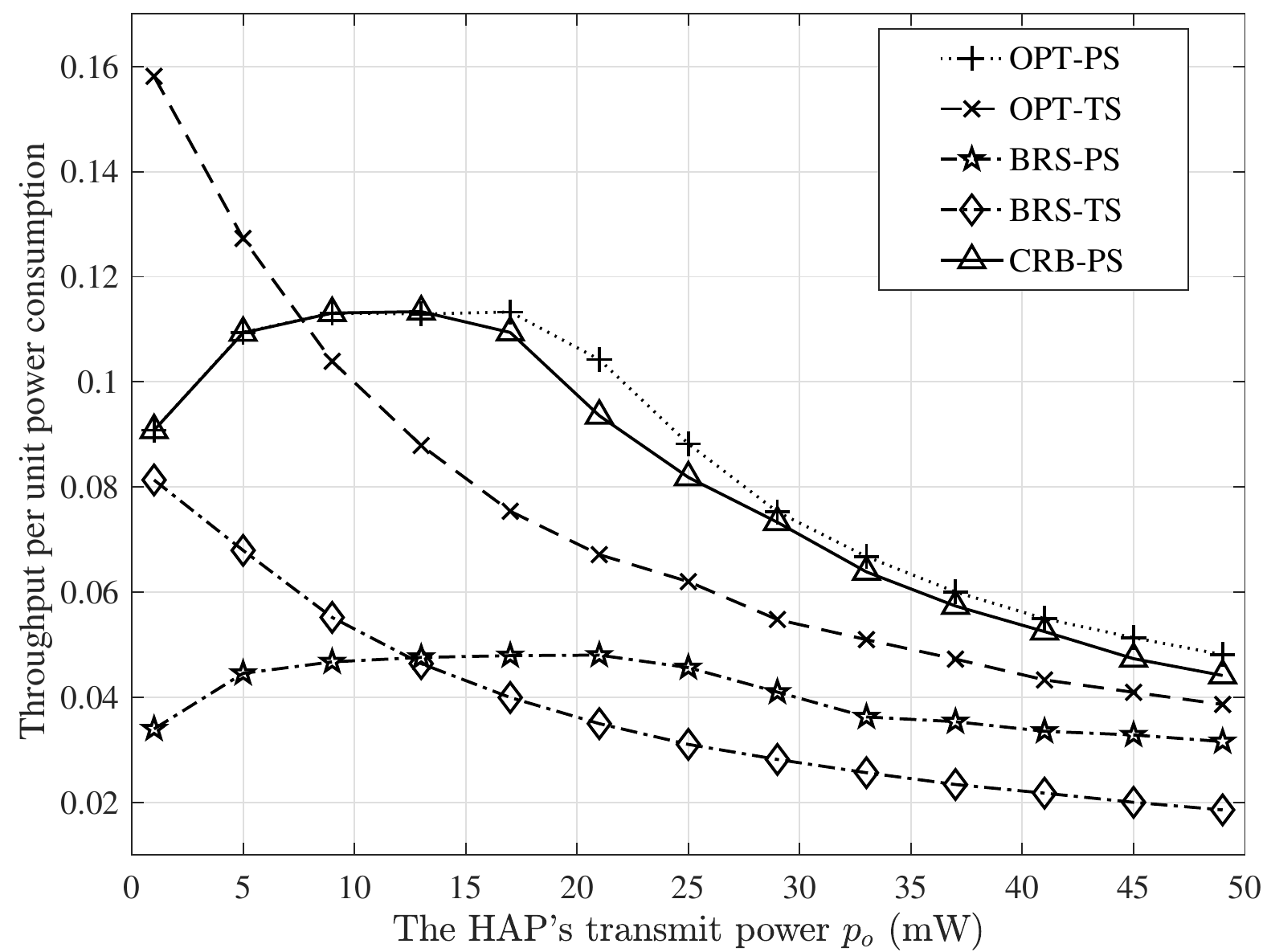}\label{fig:comp-energy-effi}}
 \caption{{Comparison of the TS and PS schemes in terms of throughput performance and energy efficiency}.}\label{fig:comp-ts-ps}
\end{figure}

\subsection{Comparison between the TS and PS Schemes}

The performance of both EH schemes is heavily affected by two conflicting parameters, i.e., the HAP's transmit power $p_{\mathrm{o}}$ and the probability limit $\zeta$ of interference violation at the CUE. Hence, in the comparison between the TS and PS schemes, we vary these two parameters and examine the throughput performance and energy efficiency. Besides, we compare our algorithm with the best relay selection (BRS) algorithm~\cite{ding16}, which selects a single relay with the best channel condition to amplify and forward the information signals. The authors in~\cite{liu2016wireless} considered a similar collaborative beamforming problem and optimized the PS ratios of multiple relays for a single-antenna transceiver pair. It also requires full CSI and centralized optimization at the transmitter. For a fair comparison, we extend the work in~\cite{liu2016wireless} to account for multi-antenna signal beamforming at the transmitter, and then apply the optimized PS ratios to our formulation in~\eqref{equ:orignal-form}. Note that all these algorithms have comparable overhead for information exchange. When there are $N$ relays, the total number of time slots required for channel estimation is $N+2$. In each time slot, either complex channel coefficients or scalar CSI metrics as in~\cite{ding16} are encoded and transmitted to the transmitter. In general, our proposed algorithms can be implemented with a comparable signaling overhead, but a higher throughput performance compared to that in~\cite{ding16} and~\cite{liu2016wireless}.

Let BRS-TS and BRS-PS denote the BRS algorithm under the TS and PS schemes, respectively. Let CRB-PS denote the collaborative relay beamforming algorithm maximizing the PS ratios in~\cite{liu2016wireless}. Fig.~\ref{fig:comp-ts-ps}(a) shows the throughput performance with different algorithms. The optimal throughputs with the TS and PS schemes, denoted by OPT-TS and OPT-PS, respectively, initially increases and then stabilizes when the HAP further increases its transmit power. It is clear that the stabilized throughput becomes smaller with a larger probability limit $\zeta$. When $p_{\mathrm{o}} \leq 6$, the CUE's interference constraint trivially holds due to low transmit power at the HAP and the relays. The energy available at the relays becomes the throughput bottleneck. Hence, we observe that the optimal throughput with the OPT-PS scheme shows no significant difference with different probability limit $\zeta$. When $p_{\mathrm{o}}>12$, the throughput of the OPT-PS scheme increases slowly, which means that the CUE's interference requirement comes into effect and restricts the DUE's throughput improvement. Besides, both OPT-TS and OPT-PS scheme show significant throughput improvement compared to the BRS-TS and BRS-PS schemes, respectively. This implies that our algorithm can better exploit the multi-relay cooperation gain. Focusing on the multi-relay schemes, we find that the OPT-PS scheme achieves nearly the same throughput with the CRB-PS scheme when the HAP's transmit power is relatively low, i.e., $p_{\mathrm{o}} < 15$, and it outperforms the CRB-PS scheme when $p_{\mathrm{o}}$ becomes large. This result shows the flexibility of the OPT-PS scheme for its energy-efficient response to the increase of power supply, via a joint control of the HAP's beamforming strategy and the relays' PS ratios.

It is interesting to observe a crossover in Fig.~\ref{fig:comp-ts-ps}(a) between the curves corresponding to the OPT-PS and OPT-TS schemes. In particular, the OPT-TS scheme outperforms the OPT-PS scheme when $p_{\mathrm{o}} < 10$, while the OPT-PS scheme achieves better throughput when $p_{\mathrm{o}} > 10$. When $p_{\mathrm{o}}$ is relatively small, the relays' power transfer and signal reception both suffer from severe power shortage. In this case, the optimization of the PS ratios hence has limited effect on improving the throughput. For the OPT-TS scheme, we can set a larger EH time to increase the energy accumulated at the relays, and hence potentially improve the throughput in information transmission. When $p_{\mathrm{o}}$ becomes large, both the OPT-TS and OPT-PS schemes can provide high data rates by the control of EH time and PS ratios, respectively. As such, the effective transmission time will dominate the throughput performance. Hence, we observe that the OPT-PS scheme performs better than the OPT-TS scheme due to its longer transmission time.

In Fig.~\ref{fig:comp-ts-ps}(b), we compare the energy efficiency of two schemes, which is defined as the throughput gained per unit power consumption at all relays. We set $\zeta=0.3$ and vary the transmit power $p_{\mathrm{o}}$ from 1 to 50. When $p_{\mathrm{o}}>20$, the energy efficiencies of both the OPT-TS and OPT-PS schemes decrease in $p_{\mathrm{o}}$. The reason is that when the relays accumulate sufficient energy, the throughput does not increase proportionally when we continue increasing the transmit power $p_{\mathrm{o}}$. Hence, the energy efficiency will correspondingly decrease as shown in Fig.~\ref{fig:comp-ts-ps}(b). When the HAP's transmit power is relatively low, i.e., $p_{\mathrm{o}} < 10$, the energy efficiency with the OPT-PS scheme grows up with the increase in $p_{\mathrm{o}}$ and reaches its maximum when $p_{\mathrm{o}}=10$. That is because, the throughput initially grows much faster than the HAP's power consumption. By contrast, for the OPT-TS scheme, the throughput grows slower than the HAP's power consumption, and thus we observe a decreasing energy efficiency in Fig.~\ref{fig:comp-ts-ps}(b). This result implies that the OPT-PS scheme responds in a more energy efficient way to the increase of the HAP's transmit power.

\begin{figure}[t]
\centering
\subfloat[Adapt PS ratios to balance power supply and demands]{\includegraphics[width=\singlesize\textwidth]{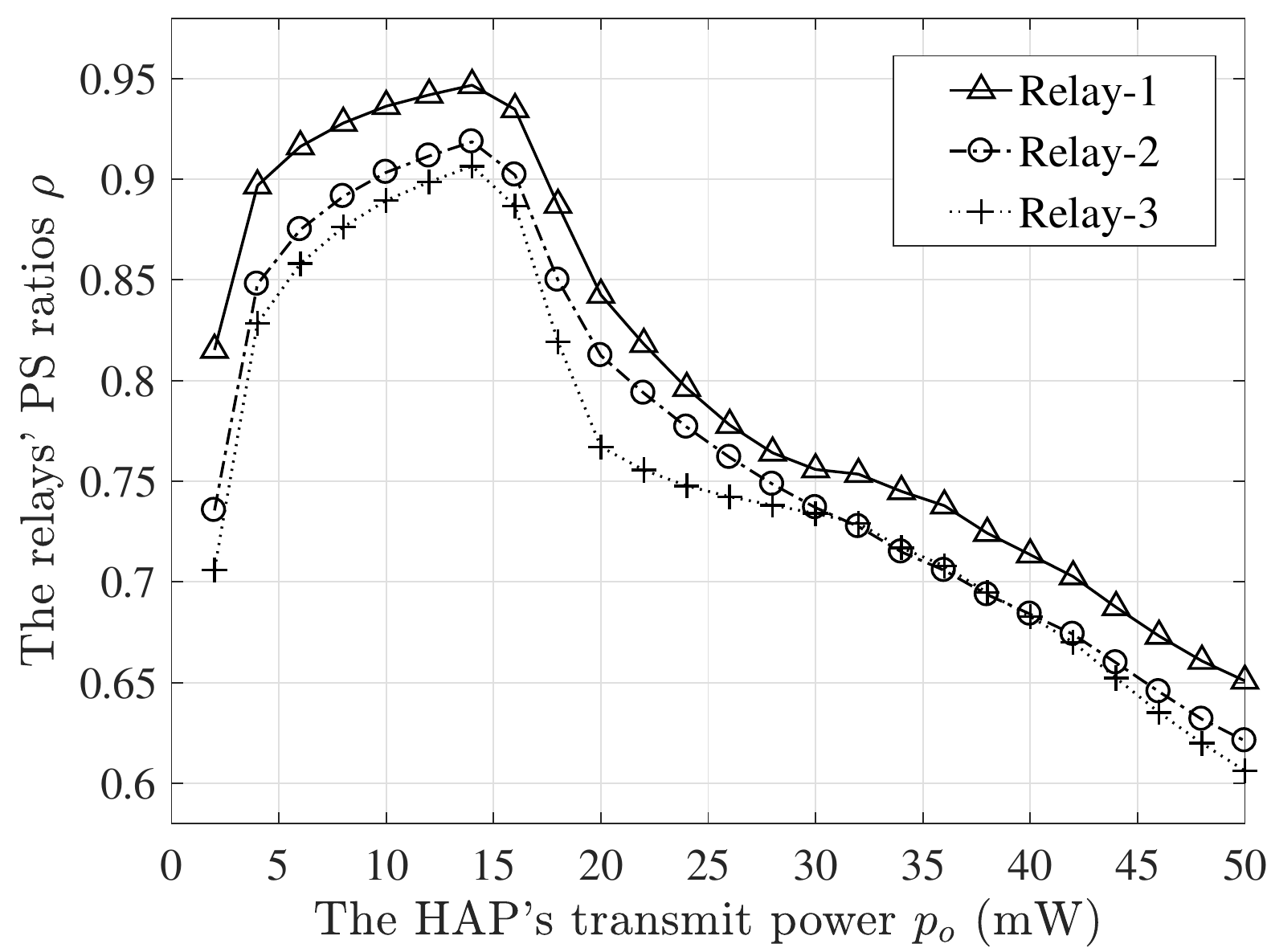}\label{fig:opt-rho}}\\
\subfloat[Biased power transfer to meet power demands]{\includegraphics[width=\singlesize\textwidth]{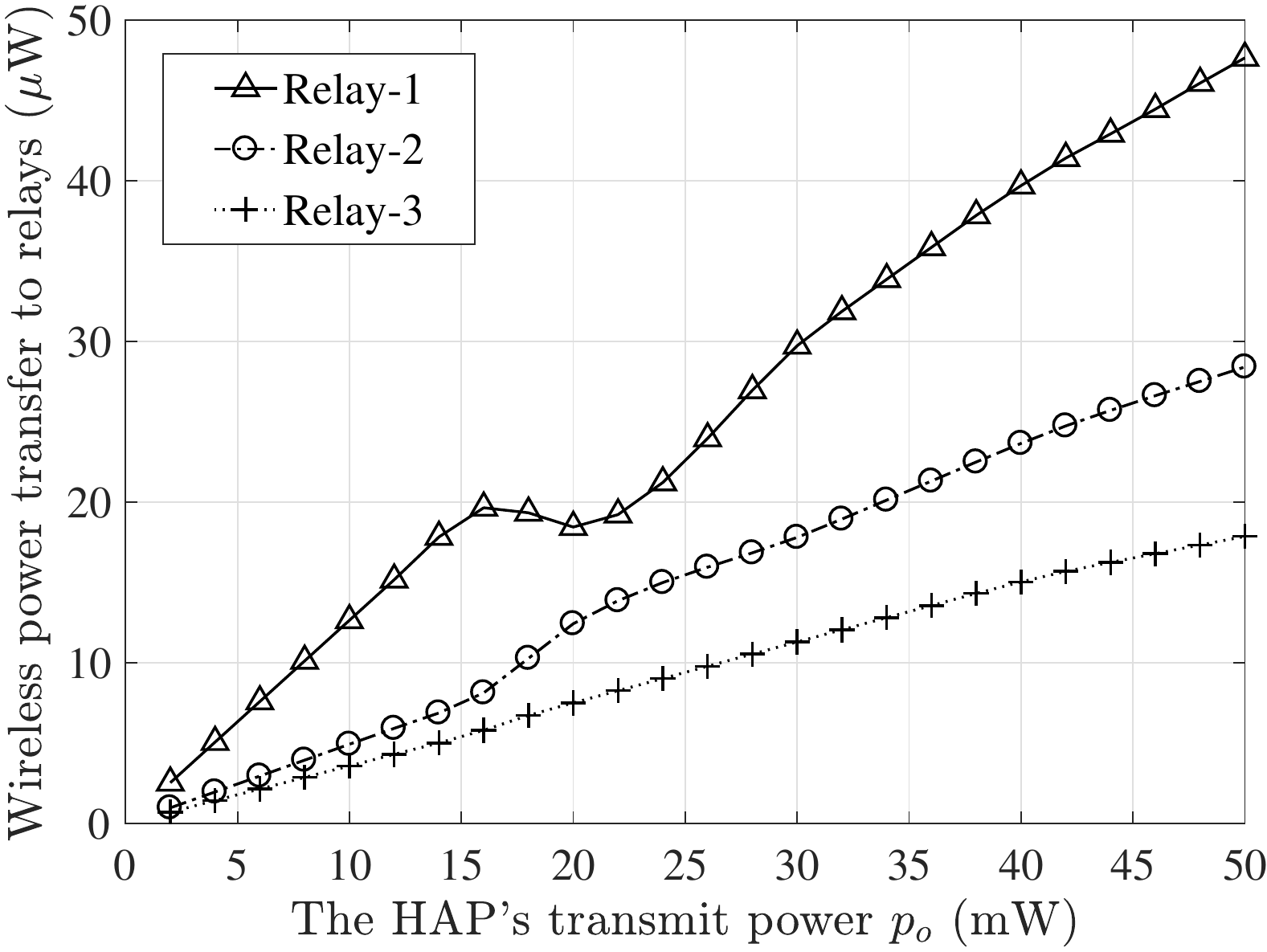}\label{fig:recv-power}}
\caption{{Joint control of the power transfer to different relays and the relays' PS ratios.}}\label{fig_control}
\end{figure}
\subsection{{Joint Power Transfer, Power Splitting, and Power Control}}\label{sec:ps-control}

The higher energy efficiency of the PS scheme can be viewed as a result from a more flexible control over the HAP's beamforming, the relays' PS ratios and transmit power. In this part, we investigate how these control variables interplay with each other. With the varying transit power $p_{\mathrm{o}}$, we show the joint control of the power transfer to the relays and the relays' PS ratios in Fig.~\ref{fig_control}. Fig.~\ref{fig_control}(a) shows the dynamics of the relays' optimal PS ratios with a fixed probability limit $\zeta=0.3$ at the CUE. We firstly find that the ordering $\rho_1^* > \rho_2^* > \rho_3^*$ generally holds and is consistent with the relays' channel conditions, i.e., $||\mathbf{f}_1||^2 > ||\mathbf{f}_2||^2 > ||\mathbf{f}_3||^2$. That is, more signal power will be allocated to energy harvester when channel condition is better. We further observe that the PS ratios will be reduced significantly when $p_{\mathrm{o}} > 16$. This is due to the fact that the relays become sufficient in energy when $p_{\mathrm{o}}$ further increases.

\begin{figure}[t]
\centering
\includegraphics[width=\singlesize\textwidth]{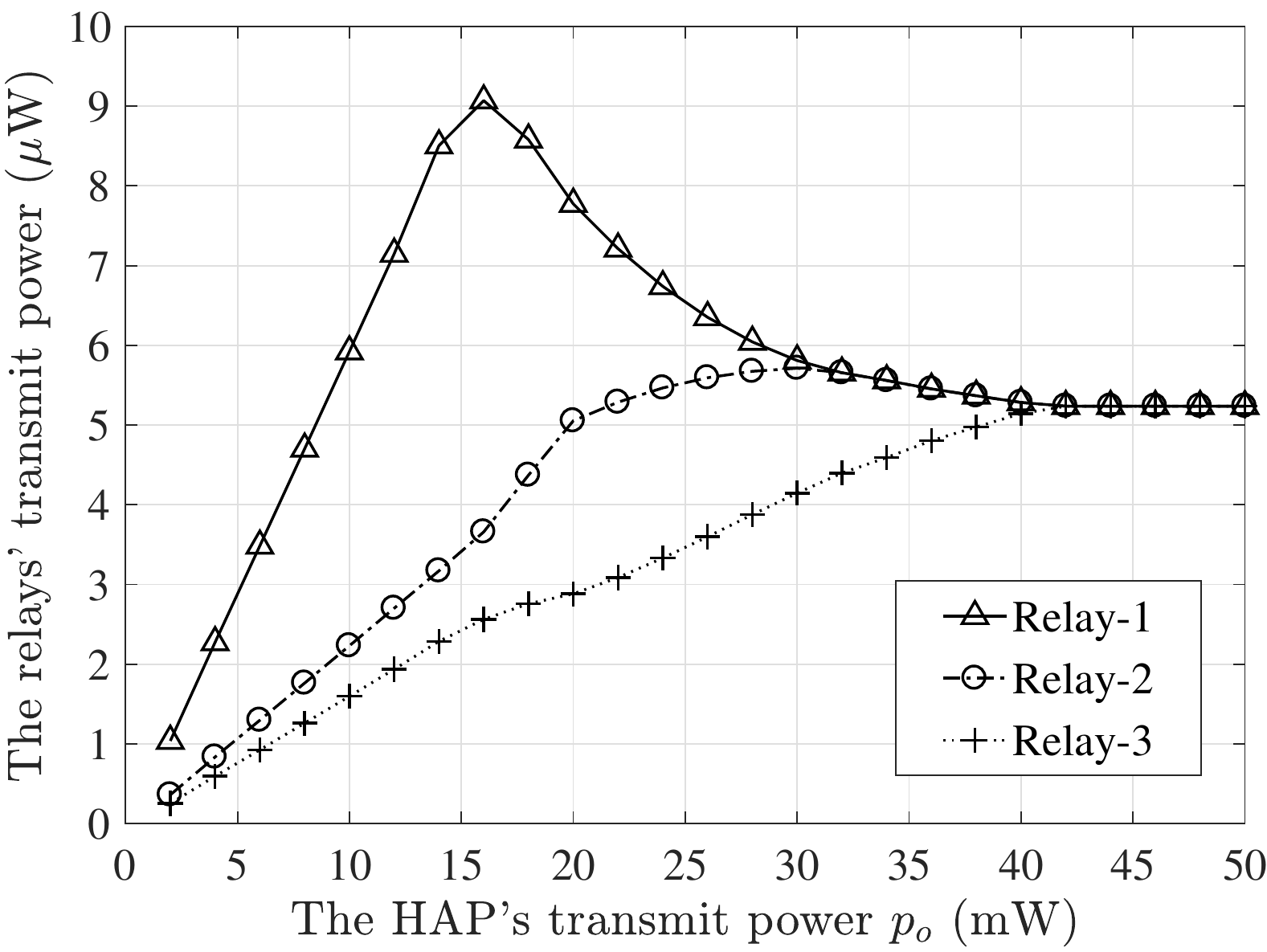}
\caption{The relays' joint power control.}\label{fig:power-control}
\end{figure}

Let $H_n = p_{\mathrm{o}} \mathbf{f}_n^H \mathbf{W} \mathbf{f}_n$ denote the power transfer to each relay $n\in\mathcal{N}$. In Fig.~\ref{fig_control}(b), we observe that the dynamics of $H_n$ show very different traces with respect to the increase of the HAP's transmit power. Specifically, we observe two turning points on each of the curves. When $p_{\mathrm{o}}<16$, the power $H_n$ increases proportionally in $p_{\mathrm{o}}$. This is reasonable when power shortage is the bottleneck of throughput improvement. The HAP will keep the same beamformer to transfer more energy to the relay with better channel condition. When $16$ $\leq p_{\mathrm{o}}\leq$ $25$, we observe a biased power transfer to different relays, due to the relays' different channel conditions and power demands. Note that relay-1 has the best channel condition. When $p_{\mathrm{o}}$ increases, relay-1 becomes energy-abundant while relay-2 and relay-3 are the energy-starving nodes that limit the throughput improvement. By updating energy beamforming strategy, the HAP can reshape the spatial distribution of RF energy, and thus adjust the energy transfer to different relays. In particular, the HAP will steer the energy beamformer towards relay-2 and relay-3 as a compensation for their worse channel conditions. As a result, we observe that the power received by relay-1 increases much slower than that of relay-2 and relay-3. When $p_{\mathrm{o}}>25$, the power $H_n$ again increases proportionally in $p_{\mathrm{o}}$. This is because the relays can harvest sufficient energy from the HAP. In this case, the HAP does not change its beamforming strategy, and the relays' transmit power control is dominated by the CUE's interference constraint.

Fig.~\ref{fig:power-control} shows the relays' power control with different $p_{\mathrm{o}}$. When $p_{\mathrm{o}} \leq 16$, the CUE's interference constraint holds and thus the relays' transmit power is linearly increasing in $p_{\mathrm{o}}$. When $p_{\mathrm{o}}\geq 16$, the CUE's interference constraint becomes active and restricts the relays' transmit power. Note that relay-1 has the highest transmit power. It introduces the highest interference to the CUE, and thus its transmit power will be significantly reduced as $p_{\mathrm{o}}$ further increases. When $p_{\mathrm{o}} \geq 40$, all relays' transmit powers stabilize at the same level, as we have assumed the same channel conditions in the second hop.

\begin{figure}[t]
\centering
\subfloat[Relay transmit power with channel uncertainty level]{\includegraphics[width=\singlesize\textwidth]{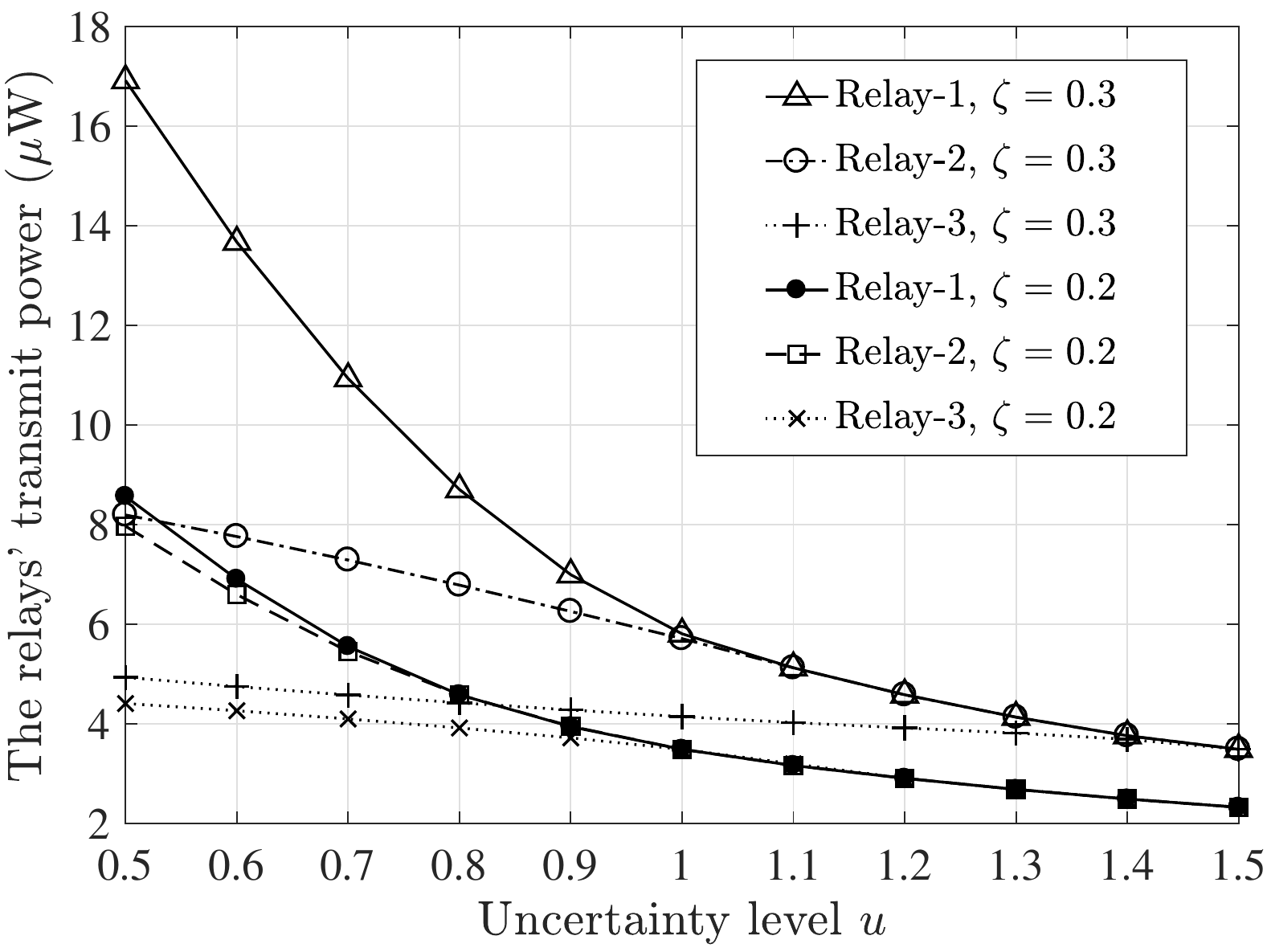}\label{fig:relay-power-uncertainty}}\\
\subfloat[PS ratios at Relays with interference violation probability]{\includegraphics[width=\singlesize\textwidth]{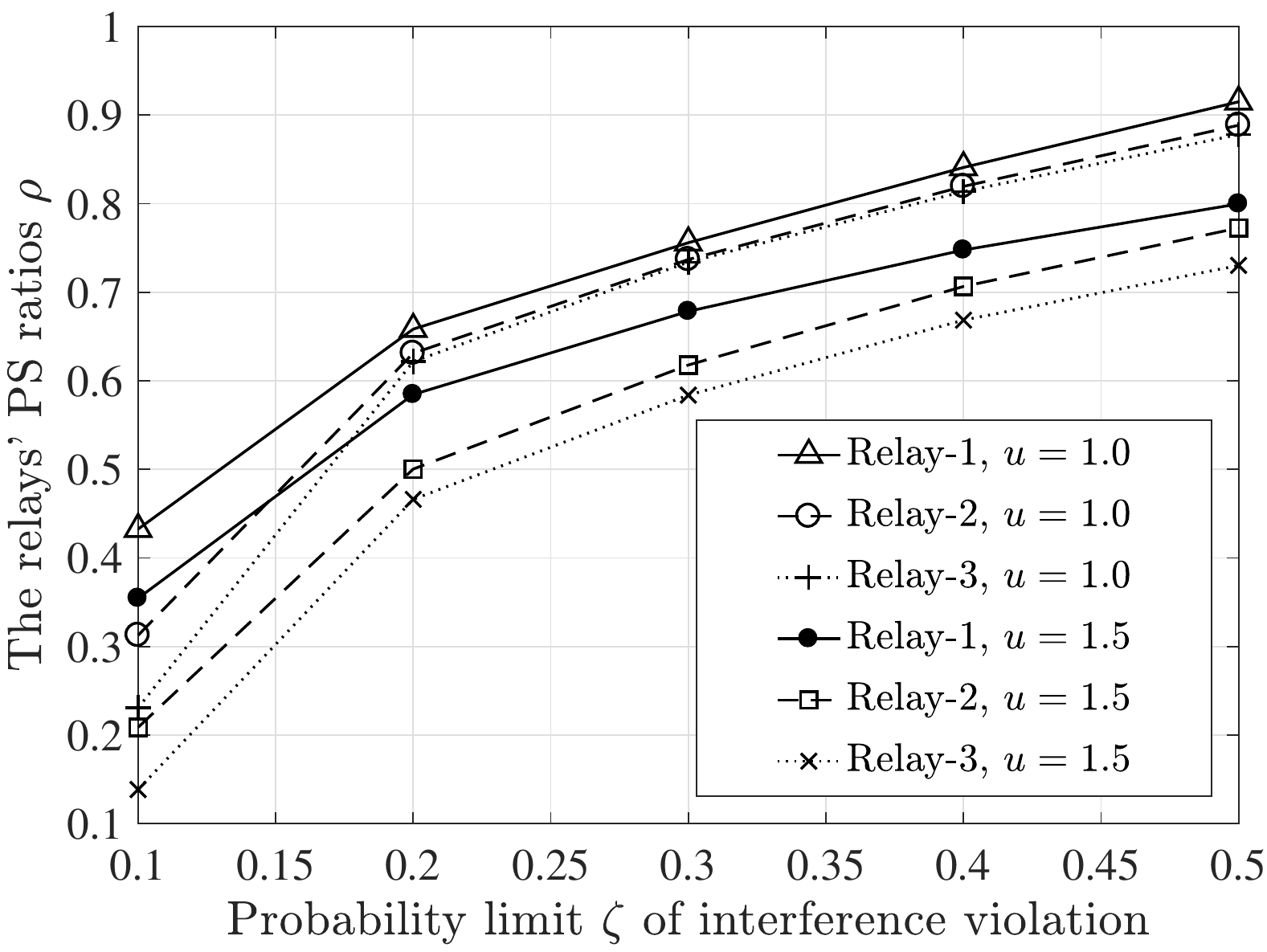}\label{fig:rho-uncertainty}}
\caption{{Relays' power and PS ratios with channel dynamics}.}\label{fig:uncertainty-strategy}
\end{figure}

\subsection{{Robustness Against Channel Uncertainty}}

We evaluate the robustness of the PS scheme against distributional channel uncertainty. By multiplying a scaling factor $u$ to ${\bf S_z}$, we emulate different levels of channel dynamics. A larger value of $u$ implies that the channel becomes fluctuating with a wider range and thus there will be interference violation with higher probability. Hence, we consider $u$ as the level of uncertainty in our experiments. Fig.~\ref{fig:uncertainty-strategy}(a) shows the relays' transmit power against different uncertainty levels. By increasing uncertainty level, we observe a decreasing transmit power at each relay as shown in Fig.~\ref{fig:uncertainty-strategy}(a). This result is intuitive as we have to suppress the relays' transmit power to ensure the same level of protection for CUEs when the channel becomes fluctuating with a wider range. We further demonstrate how the relays tune their PS ratios under different probability limit $\zeta$ as shown in Fig.~\ref{fig:uncertainty-strategy}(b). A larger $\zeta$ implies that the CUEs are more tolerant (or less sensitive) to the relays' interference. Generally, the relays' PS ratios are monotonically increasing with the increase of $\zeta$, which means that the relays need to split more signal power for information transmission when the CUEs become less sensitive to interference. Note that the PS ratios follows $\rho_1 \geq \rho_2 \geq \rho_3$. This implies that the relays with worse channel conditions need to split more power for information transmission. With wider range of the channel fluctuating, the difference between $\rho_1$ and $\rho_3$ becomes larger.

\section{Conclusion}\label{sec:conc}

In this paper, to achieve cooperation gain in a dense Device-to-Device network, multiple EH-enabled relays have been employed to improve the throughput from a multi-antenna HAP to a distant DUE receiver underlaying a cellular system. We have reviewed the typical TS and TS schemes, for the relays' power and information transfer from the HAP. A generalized optimization has been proposed to maximize the DUE's throughput subject to probabilistic interference constraints at the CUEs, by the joint design of the HAP's beamforming, the relays' energy harvesting and power control strategies. We have verified that the probabilistic interference constraint can define a normal feasible region. By further exploiting the monotonicity in the objective, we can determine a sub-optimal solution to the non-convex robust throughput maximization via successive polyblock approximation. The extensive performance evaluation has clearly shown that our algorithm design can still achieve much higher throughput than those of the existing works. This well justifies our proposed relaying scheme. We have also made thorough numerical comparison between the TS and PS schemes in terms of throughput performance and energy efficiency, and revealed that the PS scheme generally outperforms the TS scheme, due to its flexibility in the relays' local control. Such flexibility becomes trivial and lead to worse performance compared to that of the TS scheme, when the HAP's transmit power is low.

\appendix

\subsection{Proof of Proposition \ref{pro_equvi_projection}}\label{proof_equvi_projection}
We focus on the reformulation of~\eqref{con_ip_exp}. Let $e({\bf z}_m)={\bf 1}({\bf z}_m^T \text{D}({\bf c}\circ{\bf c}){\bf z}_m\geq \bar{\phi}_m)$ and ${\bf 1}(\cdot)$ be the indicator function. For $m\in\mathcal{C}$, the worst-case interference violation probability can be rewritten as $B( {\bf c} | {\bf \Sigma}_m) = \max_{\mathbb{P}\in\mathcal{P}_m} \mathbb{E}_{\mathbb{P}}\big[e({\bf z}_m)\big]$, where ${\bf \Sigma}_m$ represents the known second-order moment matrix of the channel ${\bf z}_m$. By a similar approach to that in~\cite{Globecom16:Gong}, the equivalence of $B({\bf c}|{\bf \Sigma}_m)$ is given by
\begin{subequations}\label{prob_bd_noalpha}
\begin{align}
\min_{{\bf M}_m\succeq 0,\nu_m\geq0} \,\,&\, \textbf{Tr}({\bf \Sigma}_m{\bf M}_m) \label{obj_bd_noalpha}\\
\text{s.t.} \quad\,\, &\, {\bf M }_m\succeq \left[\begin{array}{cc} \nu_m \text{D}({\bf c}\circ{\bf c}) & 0 \\ 0 & 1 - \nu_m\bar{\phi}_m \end{array}\right], \label{con1_bd_noalpha}
\end{align}
\end{subequations}
where ${\bf M}_m$ and $\nu_m$ are the dual variables. To this point, the objective function in (\ref{obj_bd_noalpha}) is linear and the constraint in (\ref{con1_bd_noalpha}) defines a linear matrix inequality. Hence, the problem in (\ref{prob_bd_noalpha}) provides a convex equivalence for $B( {\bf c}|{\bf \Sigma}_m)$. Substituting \eqref{con_ip_exp} by \eqref{prob_bd_noalpha}, we have the equivalence in problem~\eqref{prob_prjcvx}.

\subsection{Proof of Proposition~\ref{pro:normal}}\label{apd:proof-normal}
Assuming $(X^{(1)}, Y^{(1)}) \in \Omega$, for any $(X^{(2)}, Y^{(2)}) \preceq (X^{(1)}, Y^{(1)})$, we need to prove that $(X^{(2)}, Y^{(2)}) \in \Omega$. That is, we need to find a feasible solution $(\boldsymbol{\rho}^{(2)}, \mathbf{p}^{(2)},\mathbf{W}^{(2)}_{\mathrm{p}})$ satisfying the constraints listed in (\ref{equ:feasible-region}). Without loss of generality, we set $p_{\mathrm{o}} = 1$ for simplicity and let $\alpha^2 \triangleq X^{(2)} / X^{(1)}$ and $\beta^2 \triangleq Y^{(2)}/Y^{(1)}$, respectively, where $0\leq \alpha, \beta \leq 1$. Now we construct a feasible solution $(\boldsymbol{\rho}^{(2)}, \mathbf{p}^{(2)},\mathbf{W}^{(2)}_{\mathrm{p}})$ based on $(\boldsymbol{\rho}^{(1)}, \mathbf{p}^{(1)},\mathbf{W}_{\mathrm{p}}^{(1)})$. By choosing $\mathbf{W}_{\mathrm{p}}^{(2)} = \mathbf{W}_{\mathrm{p}}^{(1)}$ and $\rho^{(2)}_n$ such that $1-\rho_n^{(2)} = \beta^2 (1-\rho_n^{(1)})$, for all $n \in \mathcal{N}$, we have $1-\rho_n^{(2)} = \beta^2 \left(1-\rho_n^{(1)}\right) \leq 1-\rho_n^{(1)}$, which implies $\rho_n^{(2)}\geq\rho_n^{(1)}$.
On the other hand, by choosing $x_n^{(2)} = \alpha x_n^{(1)}$ for $n \in \mathcal{N}$, we need to prove that the above construction satisfies the constraints listed in (\ref{equ:feasible-region}). By the construction, the equality is hold. Noting that $x_n^{(2)} = \alpha x_n^{(1)}$ and $y_n^{(2)} = \beta y_n^{(1)}$, we have $\frac{p_n^{(2)}}{(1-\rho_n^{(2)}) \mathbf{f}_n^H \mathbf{W}_{\mathrm{p}}^{(2)} \mathbf{f}_n + 1} = \alpha^2\frac{p_n^{(1)}}{(1-\rho_n^{(1)}) \mathbf{f}_n^H \mathbf{W}_{\mathrm{p}}^{(1)} \mathbf{f}_n + 1}$, which implies
\begin{align*}
p_n^{(2)} ~&=~ \alpha^2 p_n^{(1)} \frac{(1-\rho_n^{(2)}) \mathbf{f}_n^H \mathbf{W}_{\mathrm{p}}^{(2)} \mathbf{f}_n + 1}{(1-\rho_n^{(1)}) \mathbf{f}_n^H \mathbf{W}_{\mathrm{p}}^{(1)} \mathbf{f}_n + 1}  \\
~&\leq~ \alpha^2 \eta \rho_n^{(1)} \mathbf{f}_n^H \mathbf{W}_{\mathrm{p}}^{(1)} \mathbf{f}_n ~ \leq ~ \eta \rho_n^{(2)} \mathbf{f}_n^H \mathbf{W}_{\mathrm{p}}^{(2)} \mathbf{f}_n.
\end{align*}
Therefore, $p_n^{(2)}$ satisfies the power constraint. A smaller transmit power also ensures the satisfaction of the interference constraint, which completes the proof that $\Omega$ is a normal set.

\subsection{Proof for Proposition~\ref{pro:optimal-C}}\label{apd:proof-optimal-C}

Let $\theta_k^*\triangleq\sqrt{Y_k / X_k}$. We will prove that given any solution to \eqref{equ:feasible-form} under two cases: (1) $\theta_k^{(1)} > \theta_k^*$ and (2) $\theta_k^{(2)} < \theta_k^*$, there always exists a solution to \eqref{equ:feasible-form} by setting $\theta_k=\theta_k^*$. By the equality constraint in \eqref{equ:equ-constr}, we can simplify \eqref{equ:A-bound} and \eqref{equ:B-bound} into a new constraint: $\lVert {\bf y}\rVert^{2} \geq \max \{\theta_k^2 \lambda_k X_k, \lambda_k Y_k\}$. Suppose that $\left(\boldsymbol{\rho}^{(i)}, {\bf p}^{(i)}, {\bf W}^{(i)}_{\mathrm{p}}\right)$ is a solution to \eqref{equ:feasible-form} when $\theta_k=\theta_k^{(i)}$ for $i=\{1,2\}$. For $\theta_k^{(1)}>\theta_k^*$, let $\alpha \triangleq \theta_k^* / \theta_k^{(1)}$ and we can choose ${\bf W}_{\mathrm{p}}^* = {\bf W}_{\mathrm{p}}^{(1)}$, ${\bf x}^* = {\bf x}^{(1)}$ and ${\bf y}^* = \alpha {\bf y}^{(1)}$. Therefore, we have
\[
|| {\bf y}^*||^{2} = \alpha^2 || {\bf y}^{(1)}||^{2} \geq \alpha^2 \big(\theta_k^{(1)}\big)^2 \lambda_k X_k = \lambda_k X_k (\theta_k^*)^2 = \lambda_k Y_k.
\]
Besides, by the construction $x_n^* = x_n^{(1)}$, we have $\frac{p_n^{*}}{(1-\rho_n^{*}) \mathbf{f}_n^H \mathbf{W}_{\mathrm{p}}^{*} \mathbf{f}_n + 1} = \frac{p_n^{(1)}}{(1-\rho_n^{(1)}) \mathbf{f}_n^H \mathbf{W}_{\mathrm{p}}^{(1)} \mathbf{f}_n + 1}$, which implies that
\[
p_n^{*} = p_n^{(1)} \frac{(1-\rho_n^{*}) \mathbf{f}_n^H \mathbf{W}_{\mathrm{p}}^{*} \mathbf{f}_n + 1}{(1-\rho_n^{(1)}) \mathbf{f}_n^H \mathbf{W}_{\mathrm{p}}^{(1)} \mathbf{f}_n + 1} \leq  \eta \rho_n^{*} \mathbf{f}_n^H \mathbf{W}_{\mathrm{p}}^{*} \mathbf{f}_n.
\]
Therefore, $p_n^{*}$ satisfies the power constraint. A smaller transmit power also ensures the satisfaction of the interference constraint. Thus, $\left(\boldsymbol{\rho}^*, {\bf p}^*,{\bf W}_{\mathrm{p}}^*\right)$ is a solution to \eqref{equ:feasible-form} when $\theta_k=\theta_k^*$.

For $\theta_k^{(2)} < \theta_k^*$, let $\beta \triangleq \theta_k^{(2)}/ \theta_k^*$ and we can choose ${\bf W}_{\mathrm{p}}^* = {\bf W}_{\mathrm{p}}^{(2)}$, ${\bf y}^* = {\bf y}^{(2)}$ and ${\bf x}^* = \beta {\bf x}^{(2)}$. Therefore $|| {\bf y}^*||^{2} = || {\bf y}^{(2)}||^{2} \geq \lambda_k Y_k = \lambda_k X_k (\theta_k^*)^2$. Besides, by the construction $x_n^* = \beta x_n^{(2)}$ and $y_n^* = y_n^{(2)}$, we have
\[
p_n^* = \beta^2 p_n^{(2)} < \eta \rho_n^{(2)} p_{\mathrm{o}} \mathbf{f}_n^H \mathbf{W}_{\mathrm{p}}^{(2)} \mathbf{f}_n = \eta \rho_n^{*} p_{\mathrm{o}} \mathbf{f}_n^H \mathbf{W}_{\mathrm{p}}^{*} \mathbf{f}_n.
\]
Till this point, we can conclude that the optimal $\theta_k$ to \eqref{equ:feasible-form} is given by $\theta_k=\sqrt{Y_k / X_k}$.

\subsection{Equivalence between problems \eqref{equ:feasible-reform-1} and \eqref{equ:final-form}}\label{proof_equvi_ps}
By the change of variable $\kappa_n = y_n^2 \geq 0$ and let $\boldsymbol{\kappa} = [\kappa_1,\ldots,\kappa_n,\ldots,\kappa_N]^T$, the quadratic objective $||{\bf y}||^2$ is transformed into a linear form $||\boldsymbol{\kappa}||_1$. To simplify the equality constraint \eqref{equ:equ-constr}, we can substitute \eqref{equ:equ-constr} into \eqref{equ_ampps} and rewrite it into an inequality constraint as follows:
\[
p_n \theta_k^2 g_n^2 \geq \left[(1-\rho_n) p_{\mathrm{o}} \mathbf{f}_n^H \mathbf{W}_{\mathrm{p}} \mathbf{f}_n + 1\right] y_n^2 \geq (\kappa_n + 1) \kappa_n,
\]
which can be further rewritten into the linear matrix inequality \eqref{con_lmi}. To decouple the PS ratio $\rho_n$ and the beamformer ${\bf W}_{\mathrm{p}}$ in \eqref{equ:power-constr}, we introduce a new matrix variable $\bar{\bf W}_{\mathrm{p}}$ such that
\begin{equation}\label{equ:new-equ-constr}
\kappa_n \leq  (1-\rho_n) p_{\mathrm{o}} \mathbf{f}_n^H \mathbf{W}_{\mathrm{p}} \mathbf{f}_n = p_{\mathrm{o}} \mathbf{f}_n^H \mathbf{W}_{\mathrm{p}} \mathbf{f}_n - p_{\mathrm{o}} \mathbf{f}_n^H \bar{\mathbf{W}}_{\mathrm{p}} \mathbf{f}_n.
\end{equation}
The first inequality in \eqref{equ:new-equ-constr} is the convex relaxation of $\kappa_n = y_n^2$ and the second equality is by construction. Note that \eqref{equ:new-equ-constr} defines a linear matrix inequality. The convex reformulation of \eqref{equ:inter-constr-form} is given by \eqref{con_prob1}-\eqref{con_prob3}, following a similar approach as that in~\cite{Globecom16:Gong}. Combining all convex equivalences into one formulation, we can obtain the equivalence in (\ref{equ:final-form}) which can be solved efficiently by well-known optimization toolboxes.

\bibliographystyle{IEEEtran}
\bibliography{02_reference}

\end{document}